\documentclass[fleqn,usenatbib]{mnras}

\usepackage{newtxtext,newtxmath}

\usepackage[T1]{fontenc}

\DeclareRobustCommand{\VAN}[3]{#2}
\let\VANthebibliography\thebibliography
\def\thebibliography{\DeclareRobustCommand{\VAN}[3]{##3}\VANthebibliography}

\usepackage{graphicx}	
\usepackage{amsmath}	

\title[FLAMINGO: Cluster selection effects]{The FLAMINGO Project: A comparison of galaxy cluster samples selected on mass, X-ray luminosity, Compton-Y parameter, or galaxy richness}

\author[Roi Kugel et al.]{
Roi Kugel,$^{1}$\thanks{E-mail: kugel@strw.leidenuniv.nl}
Joop Schaye,$^{1}$
Matthieu Schaller,$^{2,1}$
Ian G. McCarthy,$^{3}$
Joey Braspenning,$^{1}$
John C. Helly,$^{4}$\newauthor
Victor J. Forouhar Moreno,$^{1}$
Robert J. McGibbon$^{1}$
\\
$^{1}$Leiden Observatory, Leiden University, PO Box 9513, 2300 RA Leiden, the Netherlands\\
$^{2}$Lorentz Institute for Theoretical Physics, Leiden University, PO box 9506, 2300 RA Leiden, the Netherlands\\
$^{3}$Astrophysics Research Institute, Liverpool John Moores University, Liverpool L3 5RF, UK\\
$^{4}$Institute for Computational Cosmology, Department of Physics, University of Durham, South Road, Durham, DH1 3LE, UK\\
}

\date{Accepted XXX. Received YYY; in original form ZZZ}

\pubyear{2015}

\begin{document}
\label{firstpage}
\pagerange{\pageref{firstpage}--\pageref{lastpage}}
\maketitle

\begin{abstract} 
Galaxy clusters provide an avenue to expand our knowledge of cosmology and galaxy evolution. Because it is difficult to accurately measure the total mass of a large number of individual clusters, cluster samples are typically selected using an observable proxy for mass. Selection effects are therefore a key problem in understanding galaxy cluster statistics. We make use of the $(2.8~\rm{Gpc})^3$ FLAMINGO hydrodynamical simulation to investigate how selection based on X-ray luminosity, thermal Sunyaev-Zeldovich effect or galaxy richness influences the halo mass distribution. We define our selection cuts based on the median value of the observable at a fixed mass and compare the resulting samples to a mass-selected sample. We find that all samples are skewed towards lower mass haloes. For X-ray luminosity and richness cuts below a critical value, scatter dominates over the trend with mass and the median mass becomes biased increasingly low with respect to a mass-selected sample. At $z\leq0.5$, observable cuts corresponding to median halo masses between $M_\text{500c}=10^{14}$ and $10^{15}~\rm{M_{\odot}}$ give nearly unbiased median masses for all selection methods, but X-ray selection results in biased medians for higher masses. For cuts corresponding to median masses $<10^{14}$ at $z\leq0.5$ and for all masses at $z\geq1$, only Compton-Y selection yields nearly unbiased median masses. Importantly, even when the median mass is unbiased, the scatter is not because for each selection the sample is skewed towards lower masses than a mass-selected sample.  Each selection leads to a different bias in secondary quantities like cool-core fraction, temperature and gas fraction.
\end{abstract}

\begin{keywords}
 galaxies: clusters: general -- galaxies: clusters: intracluster medium – large-scale structure of Universe –– X-rays: galaxies: clusters
\end{keywords}



\section{Introduction}
Galaxy clusters are the largest virialized structures in the universe, and are found at the intersections of the filamentary network of the cosmic web. Following hierarchical structure formation, clusters are the last objects to form. Both the number density of clusters as a function of cluster mass, i.e. the halo mass function (HMF), and the properties of individual clusters are sensitive to the underlying cosmological model \citep[for a review see][]{ClusterReviewAll2011}.

The current standard model of cosmology involves a spatially flat universe dominated by dark energy and cold dark matter, and is denoted as $\Lambda$CDM. Recent weak lensing and distance ladder measurements have exposed tensions between the $\Lambda$CDM parameters recovered by measurements of the cosmic microwave background \citep[e.g.][]{Planck2020} and observations of the late-time universe \citep[e.g.][]{KIDS2021,DESYR3,Ries2022,HSC2023}. Galaxy clusters are an independent probe that can help further investigate these tensions.

The cluster cosmology probe that is used the most is cluster counts, which are parameterised via the HMF. The HMF gives the abundance of clusters as a function of their total mass within some 3D aperture, which is generally not directly observable. Instead, measurements are limited to indirect probes of the total (3D) mass of the cluster, and selection effects have to be accounted for.  Clusters are selected based on their Sunyaev-Zeldovich signal (SZ) \citep[e.g.][]{PlanckClustercosmo2016,SPTclustercosmo2019,Bleem2024}, X-ray luminosity \citep[e.g.][]{XXL2018,eFEDSclustercosmo2023,Ghirardini2024}, galaxy richness \citep[e.g.][]{Redmapper2014,Redmapper2016,Reddragon2022} combined with weak lensing signal \citep[e.g.][]{SDSSClustercosmo2019}. Future data releases of eRosita, and upcoming weak lensing missions like Euclid \citep{Euclidclusters2022} will lead to an enormous increase in the number of detected clusters. With increased statistics the cosmology constraints will become much tighter.

Systematic differences between X-ray- and SZ-selected samples are well documented observationally. \citet{Lovisari2017} reports finding an excess of disturbed clusters in SZ selected samples with respect to X-ray-selected samples. Additionally, \citet{Andrade2017} and \citet{Rossetti2017} report a larger fraction of cool-core objects for X-ray selection compared to SZ selection. \cite{Chon2017} argue that many of the differences in results for different samples are due to the difference between flux- and volume-limited selections rather than the specific selection observable used. There are quite a few comparisons of X-ray vs richness selected samples. In general, good agreement is found for the mass-luminosity relation, luminosity-richness relation, disturbed fraction and merger fraction when comparing X-ray and richness selected samples \citep[e.g.][]{RamosCeja2022,Upsdell2023}, but slight differences might exist in the X-ray luminosity-temperature relation \citep{Giles2022}. Additionally, \citet{Ota2023} find that clusters selected on having a high galaxy richness have a smaller fraction of relaxed clusters compared to X-ray selected samples. In general, galaxy richness selected samples contain a much larger number of clusters than X-ray selected samples. \citet{Grandis2021} find that this might originate from the fact that the contamination in richness selected samples increases towards lower values of richness. In a comparison between weak lensing shear selected sources and X-ray selected sources by \citet{Willis2021}, a large fraction of the sources is not matched between the catalogues. This is partially due to projection effects boosting the shear, but also because extended high flux sources were missed due to the morphological selection criteria and the XMM beam. \citet{marini2024detecting} show using mock observations that eROSITA is unable to find all group size objects, with a bias towards detecting objects with a high relative gas fraction. These differences show that every selection has a unique selection function. 

Understanding the influence of selection effects on derived cluster properties is important beyond cluster cosmology. For example, scaling relations for clusters, in particular their baryon and gas content, provide constraints on how baryons impact the matter power spectrum \citep[e.g.][]{Baryonification2019,Baryonifsim2021,Baryonification2_2021,Stijn2021,Jaime2023}. Current measurements of the gas fraction in clusters \citep[as collated by][]{Kugel2023} indicate that selection effects start to dominate for haloes around the group mass of $M_{\rm 500c} \lesssim 10^{13.5}~\rm{M}_{\odot}$.

The careful modelling of selection functions is one of the main ingredients of cosmological inference with cluster counts. As shown by \cite{Mantz2019} a good grip on both the selection criteria and the mass-observable relation is necessary. In order to do unbiased cosmology inference, proper modelling of the observable relations and the effects of the selection procedures is key \citep{Angulo2012}. Power law relations with scatter are commonly assumed to relate observables to masses \citep[e.g.][]{Rozo2014,Evrard2014,PlanckClustercosmo2016,XXL2018,Selec2020,Chaubal2022}. Especially for observations probing lower masses, these assumptions might break down, and lead to biased results. Recent cosmological inferences often combine X-ray or SZ selection with scaling relations based on lensing or richness \citep{Bocquet2024,Clerc2024,Ghirardini2024}, leading to additional complexity when modelling the selection function. A solution is to predict quantities that are directly observable. One candidate is the aperture lensing mass, as discussed by \citet{Stijn2022a,Stijn2022b}. Similarly, \citet{andreon2024observed} introduce the X-ray surface brightness within 300~kpc as a promising candidate that reduces observational biases when compared with X-ray, SZ or galaxy richness selected samples.

Cosmological constraints are typically inferred by comparing observed cluster counts to results based on (emulators of) the HMF of dark matter only simulations \citep[e.g.][]{Tinker2008,MiraTitanHMF2020}. However, baryonic physics can lead to biases \citep[e.g.][]{Stijn2021}. Additionally, dark matter only simulations cannot self-consistently model the gas that is needed to predict X-ray and SZ observables. As hydrodynamical simulations are computationally more expensive than dark matter only simulations, some of the state-of-the-art simulations like EAGLE \citep{Eagle2015}, Horizon-AGN \citep{HorizonAGN2017}, IllustrisTNG \citep{TNG2018} and Simba \citep{Simba2019} do not sample volumes sufficiently large to contain a representative sample of clusters. Simulations like BAHAMAS \citep{Bahamas2017}, MilleniumTNG \citep{MTNG2022} have volumes large enough to investigate typical clusters at low redshift, but for converged statistics for the halo mass distributions even larger volumes are needed. While the lowest-resolution simulations of the Magneticum suite \citep{Magneticum2014} have large volumes, so far only BAHAMAS uses subgrid models that have been calibrated to reproduce the gas fractions of clusters. Cosmological hydrodynamics simulations can be extended to the cluster range by making use of zoom-in simulations \citep[e.g.][]{MACSIS2017,Hydranga2017,Hahn2017,Threehundred2018,Rhapsody2023,ClusterTNG2023}. While zooms enable simulating samples of massive clusters without the need to model very large volumes, they require selecting a sample from a large volume dark matter only simulation. Because a volume-limited sampled cannot be constructed from zooms, they cannot yield an unbiased study of selection effects.

For this work we make use of the FLAMINGO simulations \citep{FLAMINGOmain,Kugel2023}. FLAMINGO is a suite of large-volume cosmological hydrodynamical simulations in box-sizes with side-lengths of $1.0$ and $2.8~\rm{Gpc}$. At a resolution of $m_{\rm gas}=1.07\times10^{9}~\rm{M}_{\odot}$, using $5040^3$ gas particles, the $(2.8~\rm{Gpc})^3$ FLAMINGO box is the largest cosmological hydrodynamics simulation evolved to $z=0$. Additionally, FLAMINGO includes models that vary the resolution, cosmology, and feedback strength in boxes of $(1.0~\rm{Gpc})^3$. The cluster gas fractions and stellar mass function of the fiducial and feedback variations are calibrated to shifted observations.

The FLAMINGO simulations have been shown to be in good agreement with observations of hot gas in groups and clusters \citep{FLAMINGOmain,Braspenning2023}. In particular, \citet{Braspenning2023} find that the X-ray luminosity, temperature and thermal SZ scaling relations are in good agreement with the data at all redshifts. The thermodynamic profiles also agree well with the observations, although the metallicities are too high in cluster cores. \citet{Braspenning2023} also find that the cool-core fractions are difficult to compare with observations, as they are very dependent on the measure used, and typically based on the properties of the gas at radii near or below our resolution limit, but that they are in agreement with other simulation projects. The cool-core fractions vary more strongly between the FLAMINGO feedback variations than is the case for the scaling relations and the outer thermodynamic profiles.

The simulation's very large volume, containing 461 (4100) clusters of mass $M_{\rm 500c}$\footnote{$M_{\rm 500c}$ is the mass enclosed by a sphere with radius $R_{\rm 500c}$, which is defined as the radius of a sphere centered on a halo within which the average density is 500 times the critical density.}$> 10^{15}~\rm{M}_{\odot} (5\times10^{14}~\rm{M}_{\odot})$ at $z=0$, the agreement with cluster observations, as well as the availability of convergence tests and model variations, make FLAMINGO ideal for investigating the impact of selection effects on cluster counts.

We will compare selections based on X-ray luminosity, integrated thermal SZ effect and galaxy richness. We will contrast these selections with mass-selected samples for different redshifts. We will perform all these selections on theoretical quantities, without applying any other observational biases, projection effects, or noise. Our results are thus for a best-case scenario as selection effects are likely to become stronger when the sample selection is forward modelled using virtual observations. We choose to limit this work to theoretical quantities to increase the interpretability and because further steps towards forward modelling require choices that are survey specific, which will make the results less general. This is also why we leave an investigation of lensing masses to future work, as lensing only works in projection and requires the specification of a survey-specific redshift distribution of lenses galaxies. In future work we plan to model selection effects for specific observables and surveys by creating virtual observations based on FLAMINGO's lightcone output.

This paper is structured as follows: In Section~\ref{sec:methods} we discuss the FLAMINGO simulations, the quantities we select on and our definition of the sample mass bias, in Section~\ref{sec:results} we present our results and we conclude and summarise our findings in Section~\ref{sec:conc}.

\section{Methods}\label{sec:methods}
In this section we describe the methods and data used. We discuss the FLAMINGO simulations and how we obtain halo catalogues in Section~\ref{sec:FLAM}. The definitions used for the different quantities are described in Section~\ref{sec:quants} and the metrics with which we quantify the quality of the selections are described in Section~\ref{sec:selecs}.
\subsection{FLAMINGO}\label{sec:FLAM}
This work makes use of the FLAMINGO simulations, described in detail by \citet{FLAMINGOmain}. FLAMINGO (Full-hydro large-scale structure simulations with all-sky mapping for the interpretation of next generation observations) is a suite of cosmological hydrodynamics simulations in large volumes with variations in baryonic feedback, cosmology, box size and resolution. In this work we make use of the simulations run at intermediate resolution ($m_{\rm gas}=1.07\times10^{9}~\rm{M}_{\odot}$) in a volume of $(2.8~\text{Gpc})^3$ which consist of $2\times5040^3$ gas and dark matter particles, and $2800^3$ neutrino particles. The full output consists of 79 snapshots, of which we will use the snapshots at $z=[0,0.3,0.5,1.0,2.0]$.

The FLAMINGO simulations use the open source code \textsc{Swift} \citep{SWIFT2023}. The simulations make use of the \textsc{SPHENIX} SPH scheme \citep{Sphenix2022} with a \citet{wendland1995} $C^2$ kernel. Neutrinos are simulated using the $\delta f$ method \citep{Deltaf2021}. The ICs are generated using a modified version of \textsc{Monofonic} \citep{Monofonic2021,ElbersIcs2022}. The simulations use the `3x2pt + all external constraints' cosmology from the dark energy survey year 3 results of \cite{DESYR3} $(\Omega_{\rm m} = 0.306, \ \Omega_{\rm b} = 0.0486, \ \sigma_8 = 0.807, \ {\rm H}_{0} = 68.1, \ n_{\rm s} = 0.967)$. Simulations with different cosmologies are available but not used in this work.

FLAMINGO includes subgrid models for element-by-element radiative cooling and heating \citep{Ploeckinger2020}, star formation \citep{SchayeDV2008}, stellar mass loss \citep{Wiersma2009chemistry,Eagle2015}, feedback energy from supernova \citep{DVSchaye2008kin,Chaikin2022,Chaikin2022b}, seeding and growth of black holes, and feedback from active galactic nuclei \citep{Springel2005a,BoothSchaye2009,bahe2021}. The fiducial models use a thermal model for AGN \citep{BoothSchaye2009}, but we have two variations that use kinetic jets \citep{Husko2022} \citep[for a detailed description see][]{FLAMINGOmain}. As for BAHAMAS, the important simulation parameters are set to match the observed $z=0$ galaxy stellar mass function \citep{Driver2022} and a compilation of data of gas fractions in clusters \citep{Kugel2023}. Unique to the FLAMINGO simulations is the method used to calibrate the subgrid physics. For FLAMINGO these parameters are fit to the observations by making use of emulators, as described by \citet{Kugel2023}. This procedure is also used to constrain a set of feedback variations that skirt error bars on the calibration data. The variations are denoted by the change in the observations they are matched to. "fgas$\pm N\sigma$" denotes runs where the gas fraction is shifted up or down by $N\sigma$, "M*$-\sigma$" denotes runs where the stellar mass function is shifted to lower masses by $1\sigma$ and "Jet" denotes runs where AGN feedback is implemented in the form of kinetic jets instead of thermally-driven winds.

We identify cosmic structure using a recently updated version (see Forouhar Moreno et al. in prep) of the Hierarchical Bound Tracing algorithm \citep[HBT+, ][]{HBT2017}, which leverages hierarchical structure formation to identify substructures more robustly than traditional halo finders. In short, it identifies structures as they form in isolation, by subjecting particles within spatial friends-of-friends (FOF) groups to an iterative unbinding procedure. The particles associated to these self-bound objects are tracked across outputs to provide a set of candidate substructures at later times. This allows the identification of satellites, as the particle memberships are retained once they have been accreted by the FOF of a more massive halo. Finally, each candidate substructure is subject to additional self-boundness and phase-space checks to decide whether it is still resolved, or if it has merged or disrupted.
The HBT+ catalogue is further processed by the Spherical Overdensity and Aperture Processor
(SOAP\footnote{\url{https://github.com/SWIFTSIM/SOAP}}; McGibbon et al. in prep) , which computes a large selection of halo properties in a
range of apertures. For this work we use properties inside $R_{\rm 500c}$, which is defined as the radius within which the enclosed density is 500 times the critical density. $R_{\rm 500c}$ defines the mass $M_{\rm 500c}$ which is defined as the mass within $R_{\rm 500c}$. Because observational studies of clusters focus on centrals, we consider only central galaxies, as identified by VR.
\subsection{Observables used for selection}\label{sec:quants}
The X-ray luminosity within $R_{\rm 500c}$ is defined as the intrinsic luminosity within the Rosat 0.5-2.0 keV broad band in the observer frame. This excludes star forming gas and gas at low temperatures ($T<10^{5}~\rm{K}$). We do not attempt to exclude satellites and sum over all particles within $R_{\rm 500c}$. The X-ray luminosity of each particle is computed by interpolating in redshift, density, temperature and individual element abundances, based on output from the photo-ionisation spectral synthesis code Cloudy \citep{CLOUDY2017}. A detailed description is given by \citet{Braspenning2023}. Because the luminosities are measured in the observer frame, different parts of the rest-frame X-ray spectra will fall in the band at different redshifts.

We measure the thermal SZ Compton-Y in an aperture of $5\times R_{\rm 500c}$ as done in \citet{PLanckSZ2016}, but in Appendix~\ref{sec:cyr500} we show some of the results also for an aperture of $R_{\rm 500c}$. Compton-Y is computed by summing over the Compton-Y contribution from each individual gas particle, $y_i$, which is stored in the snapshots. The contribution of the individual particles is computed at run-time following
\begin{equation}
    y_i = \frac{\sigma_{\rm T}}{m_{\rm e} c^2} n_{\text{e},i} k_{\rm B} T_{\text{e},i} \frac{m_i}{\rho{}_i},
\end{equation}
where $\sigma_{\rm T}$ is the Thomson cross section, $m_{\rm e}$ is the electron mass, $c$ is the speed of light, $k_{\rm B}$ is the Boltzmann constant,  $n_{\text{e},i}$ is the electron number density, $T_{\text{e},i}$ is the electron temperature $m_i$ is the mass and $\rho_i$ is the density of the particle with index $i$. The electron number density and temperature are obtained from the cooling tables. Selections based on the integrated Compton-Y are referred to as SZ-selections.

For both the X-ray luminosity and Compton-Y signal we exclude particles that in which AGN feedback energy has recently been directly deposited. This can affect the X-ray luminosity, particularly for outlier haloes with a high luminosity, but has a negligible effect on Compton-Y. AGN feedback in the fiducial FLAMINGO simulations is implemented thermally, heating a single particle to a high temperature. Particles that are heated tend to be close to the core of the halo and can have very high densities. This can lead to single particles having an unrealistically large contribution to the total X-ray luminosity and Compton-Y signal of the halo, potentially dominating over the rest of the halo, which would be unphysical. To avoid this we ignore the contribution to the X-ray luminosity and Compton-Y signal of particles that have been heated in the last $15~\rm{Myr}$ and that have a temperature in the range
\begin{equation}
    10^{-1}\Delta T_{\rm AGN} \leq T_{i} \leq 10^{0.3}\Delta T_{\rm AGN},
\end{equation}
where $T_{i}$ is the temperature of the particle and $\Delta T_{\rm AGN}$ is the change in temperature when a particles is heated by a black hole, which has a value of $10^{7.78}~\rm{K}$ for the fiducial FLAMINGO model.

We define richness by counting the number of satellite galaxies above a mass threshold. Richness is defined as 
\begin{equation}
    \lambda = N_{\rm sats}(M_{*} > 10^{10.046}~\text{M}_{\odot}, r<R_{\rm 200c} ) + 1,
\end{equation}
where $M_{*}$ is the stellar mass within a $50$~proper~kpc spherical aperture and $r$ is the spherical radius from the centre of the cluster. These mass and radial limits were chosen to be similar to the cuts used for Redmapper \citep{Redmapper2014}. The mass limit is obtained from the fact that Redmapper uses a cut of $0.2L_{*}$, where $L_{*}$ is the luminosity at the knee of a Schecter fit to the luminosity function. We convert this to $0.2 M_{*}$ and use the mass at the knee from the stellar mass function of \citet{Driver2022}, which FLAMINGO is calibrated to match. The Redmapper radial cut is a function of richness, and is optimised as part of the richness finding process. We instead opt for $R_{\rm 200c}$. This gives us the scaling of the radius with halo mass that is implicit in the Redmapper radial cut, but with a pre-defined radius for each halo. We pick $R_{\rm 200c}$ over $R_{\rm 500c}$ as the satellites in the interior of the clusters are more likely to be affected by resolution-dependent tidal disruption, and a larger radius leads to better convergence. For the values of richness that we recover $R_{\rm 200c}$ is usually a factor of a few larger than the scale cut used for Redmapper. As we do not fully forward model Redmapper, we choose to use a larger 3D volume instead of a cylinder as this leads to a more well defined sample. Qualitatively the differences between a 3D sphere and a 2D projection will be small without forward modelling. The FLAMINGO simulations are calibrated to reproduce the galaxy mass function down to a stellar mass of $10^{9.9}$~M$_{\odot}$. We wish to ensure that, on average, haloes down to $M_{\rm \rm 500c}=10^{13}~\rm{M}_{\odot}$ still have more than one satellite above this mass, as a selection based on a richness of one returns all haloes. Note that Redmapper itself makes a probabilistic prediction for the number of satellites, and is hence not as affected by discreteness effects at low galaxy richness, though it will still be affected by small-number statistics for individual sources.

\subsection{Sample Selection}\label{sec:selecs}
A selection based on observable $A$ is defined as the set of haloes that have $A>A_{\rm C}$, where $A_{\rm C}$ is the selection limit. In order to compare selections based on different observables, we find the corresponding selection limits by taking the median of each observable at a fixed halo mass. In the case of an ideal scaling relation without scatter, such a selection would be equivalent to a mass selection. To compute this median value, we select haloes in a mass bin of 0.1~dex centred around the chosen mass limit. We then compute the median X-ray luminosity, thermal SZ signal or richness for these haloes. The cut, $A_{\rm C}$, is defined as
\begin{equation} \label{eq:cut_eq}
    A_{\rm C}(M_\text{C}) = \text{median}\left[A\left(10^{-0.05}M_{\rm C}<M_{\rm 500c}<10^{0.05}M_{\rm C}\right)\right],
\end{equation}
where $M_{\rm C}$ is the target mass cut. By comparing sample selections $A>A_\text{C}(M_\text{C})$ using the same target mass cut $M_\text{C}$, we can investigate how selections based on different observables deviate from the ideal case where $A$ is exactly proportional to $M_{\rm 500c}$ with no scatter.

Cluster count studies relate the counts in a sample to the HMF. To investigate how much the sample deviates from a mass-selected sample, we define the sample mass bias factor
\begin{equation} \label{eq:samplebias}
    b_{M_{\rm 500c}}(a,M_\text{C}) = \frac{\text{median}\left(M_{\rm 500c}|A>a\right)}{\text{median}\left(M_{\rm 500c}|M_{\rm 500c}>M_\text{C}\right)} - 1.
\end{equation}

Hence, $b_{M_{\rm 500c}}(a,M_\text{C})$ indicates the bias in the median $M_{\rm 500c}$ of the sample $A>a$ compared to a sample for which  $M_\text{500c} > M_\text{C}$. A bias of zero indicates an unbiased median mass. A negative (positive) bias indicates the median mass in the sample is lower (higher) than for the mass-selected sample. The bias factors for percentiles other than the median are defined analogously. Note that for the special case $a=A_{\rm C}$ the bias is only a function of $M_{\rm C}$. The bias has to be calculated separately for each redshift. By defining the sample mass bias in this way, we can quantify how much a cluster sample selected by a simple cut based on the value of an observable is influenced by lower-mass haloes that up-scatter into the sample. By investigating the bias in percentiles lower than the median, we will further quantify the level of contamination by lower mass haloes in the sample. We choose the 5th percentile as it strikes a good balance between probing the lower mass tail of each sample without being too influenced by small number statistics. The qualitative results are insensitive to the percentile picked, but in general a lower percentile that is further from the median leads to a larger value for the sample mass bias.

\begin{table}
    \centering
        \caption{Values for the fits to Eq.~\ref{eq:fiteq} at $z=0.3$. Top four rows are for X-ray luminosity, middle four for integrated Compton-Y and bottom for for galaxy richness. Note that for richness we only fit a regular lognormal, so we do not the include the power law parameters. Fits for other redshifts can be found in Appendix~\ref{sec:morez}.}
\begin{tabular}{c|c|l|l|l|l|l}
\hline
$a$ & $M_{\rm 500c}[\rm{M}_{\odot}]$ & $A$ & $\mu$ & $\sigma$ & $\log_{10}a_t$ & $\alpha$ \\
\hline
X-ray & $10^{13.0}~\rm{M}_{\odot}$ & $1.888\times10^{-3}$ & 41.1 & 0.35 & 41.6 & 1.90 \\
X-ray & $10^{13.5}~\rm{M}_{\odot}$ & $2.875\times10^{-3}$ & 42.2 & 0.23 & 42.6 & 3.67 \\
X-ray & $10^{14.0}~\rm{M}_{\odot}$ & $4.027\times10^{-3}$ & 43.2 & 0.17 & 43.5 & 3.96 \\
X-ray & $10^{14.5}~\rm{M}_{\odot}$ & $4.779\times10^{-3}$ & 44.0 & 0.14 & 44.9 & 6.50 \\
SZ & $10^{13.0}~\rm{M}_{\odot}$ & $3.882\times10^{-3}$ & -7.05 & 0.23 & -6.92 & 2.09 \\
SZ & $10^{13.5}~\rm{M}_{\odot}$ & $5.464\times10^{-3}$ & -6.09 & 0.17 & -5.98 & 3.27 \\
SZ & $10^{14.0}~\rm{M}_{\odot}$ & $7.010\times10^{-3}$ & -5.25 & 0.13 & -5.17 & 4.25 \\
SZ & $10^{14.5}~\rm{M}_{\odot}$ & $8.185\times10^{-3}$ & -4.46 & 0.12 & -4.31 & 5.04 \\
$\lambda$ & $10^{13.0}~\rm{M}_{\odot}$ & $2.729\times10^{-2}$ & 0.39 & 0.34 & - & - \\
$\lambda$ & $10^{13.5}~\rm{M}_{\odot}$ & $2.391\times10^{-2}$ & 0.91 & 0.24 & - & - \\
$\lambda$ & $10^{14.0}~\rm{M}_{\odot}$ & $3.511\times10^{-2}$ & 1.31 & 0.17 & - & - \\
$\lambda$ & $10^{14.5}~\rm{M}_{\odot}$ & $4.534\times10^{-2}$ & 1.77 & 0.13 & - & - \\
\hline
\end{tabular}
    \label{tab:fits}
\end{table}

\begin{figure}
    \centering
    \includegraphics[width=\columnwidth]{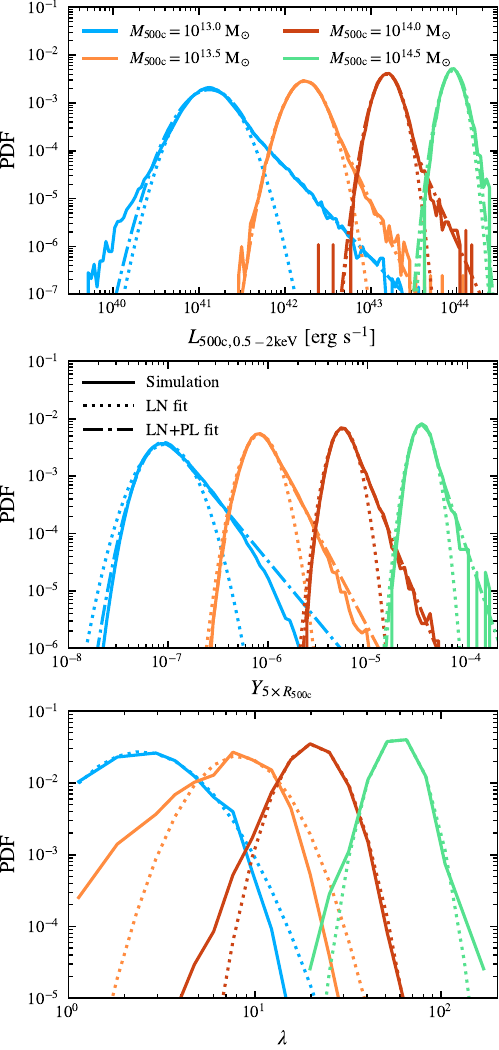}
    \caption{The distribution of X-ray luminosity (top), integrated thermal SZ Compton-Y (middle) and galaxy richness (bottom) at fixed mass at $z=0.3$. The different colours indicate different mass bins of width $\pm0.05~\rm{dex}$, around the central value. The dotted lines show the best-fitting lognormal function and, for X-ray and thermal SZ, the dot-dashed lines show the best-fitting lognormal plus power-law distributions (Eq.~\ref{eq:fiteq}). For lower masses the lognormal distributions shift to smaller values and become narrower, while  the power-law tails start at lower values and become shallower. Assuming lognormal distributions would underestimate the amount of upscatter from lower mass objects for a given cut on the value of the chosen mass proxy.}
    \label{fig:lognorm}
\end{figure}

\begin{figure*}
    \centering
    \includegraphics[width=\textwidth]{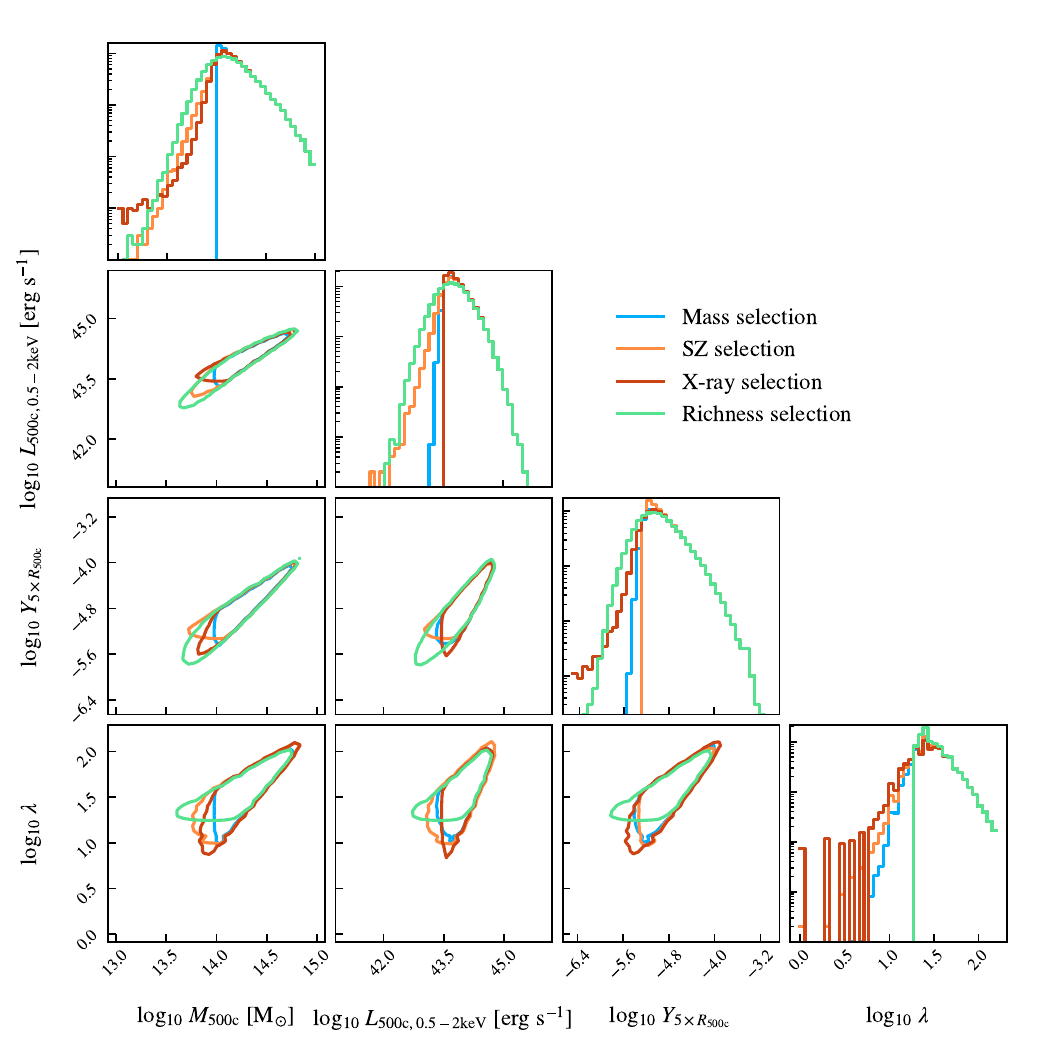}
    \caption{Corner plot showing the distribution of four different cluster properties $A$: $M_{\rm 500c}$, X-ray luminosity, Compton-Y and galaxy richness, for different selections of haloes in the 2.8~Gpc FLAMINGO fiducial volume at $z=0.3$.  Different colours correspond to samples selected based on different quantities $A$, as indicated in the legend. Each sample is defined to have $A>A_\text{C}$ where $A_\text{C} = \text{median}[A(M_{\rm 500c}=10^{14}~\rm{M}_{\odot})]$.
    The panels along the diagonal show histograms, while the off diagonal panels show two-dimensional distributions with each contour containing 95 per cent of the haloes in the sample. For each sample, the value of $A_\text{C}$ corresponds to the sharp cutoff in the histogram shown in the top panel of the column with $A$ plotted along the $x$-axis. The different samples converge for $A \gg A_\text{C}$ but there are differences for $A \lesssim A_\text{C}$.}
    \label{fig:cornerplot}
\end{figure*}

\section{Results}\label{sec:results}
In this section we compare the properties of cluster samples obtained with different selection cuts $A > A_\text{C}$, where $A$ is mass $M_\text{500c}$, X-ray luminosity $L_\text{500c,0.5-2keV}$, thermal SZ signal $Y_{5\times R_\text{500c}}$, or galaxy richness $\lambda$. In Section~\ref{sec:lognorm} we show and fit the distributions of each of the selection observables at fixed mass. We describe the general correlations between the differently selected samples for a target mass cut of $M_\text{C} = 10^{14}\,\text{M}_\odot$ in Section~\ref{sec:general_correlations}. We show different percentiles of the mass distribution as a function of $A_\text{C}$ in Section~\ref{sec:characteristic_mass}. We investigate the shift across redshift for selections based on cuts number density and observables in Sections~\ref{sec:fixed_density} and \ref{sec:fixed_mass}, respectively. In Section~\ref{sec:changing_mass} we investigate how the sample bias depends on mass and redshift. We finish by investigating how the different selections impact secondary cluster properties in Section~\ref{sec:biased_props}.

\subsection{Scatter at fixed mass}\label{sec:lognorm}
Before comparing samples defined by cuts in different observables, we will investigate the distribution of the observable mass proxies at fixed halo mass. Fig.~\ref{fig:lognorm} shows the scatter in X-ray luminosity (top panel), SZ Compton-Y (middle panel) and galaxy richness (bottom panel) in four different mass bins at $z=0.3$. The mass bins are 0.1~dex wide, $\pm0.05~\rm{dex}$ around the centre, and are centred on $\log_{10} M_\text{500c}/\text{M}_\odot = 13.0$, 13.5, 14.0, and 14.5. 

The distributions shift towards larger values for higher masses. Near their peaks, the distributions are well described by lognormal fits (dotted curves). However, the X-ray luminosity and Compton-Y distributions have tails towards higher values that deviate from the lognormal fits, skewing the distributions towards larger values. These distributions are  well fit by lognormal plus power-law functions (dot-dashed curves) parameterised as
\begin{equation}\label{eq:fiteq}
N_{\rm haloes}(a) = 
\begin{cases}
    A\exp\left(-\frac{(\log_{10}(a) - \mu)^2}{\sigma^2}\right) & a \leq a_{t}, \\
    Ba^{-\alpha} & a > a_{t},
\end{cases}
\end{equation}
where,
\begin{equation}
    B = \frac{A\exp\left(-\frac{(\log_{10}a_{t} - \mu)^2}{\sigma^2}\right)}{10^{-\alpha \log_{10}a_t}}.
\end{equation}
The best-fitting values of the free parameters $A$, $\mu$, $\sigma$, $\log_{10}a_t$ and $\alpha$, which we obtained using least squares statistics, where each bin was weighted by $1/\sqrt{N}$, can be found in Table~\ref{tab:fits}, and the result for other redshifts can be found in Appendix~\ref{sec:morez}. The general trends described below also apply to the other redshifts. 

For lower mass bins the lognormal parts become narrower, the power-law tails start closer to the peak and the slope become shallower. As a result, samples defined by a cut $A_\text{C}$ will suffer from a slight increase in upscatter from low-mass bins and this upscatter will be underestimated if the distributions are assumed to be lognormal, which is the assumption conventionally adopted in the literature. X-ray is slightly more skewed, and Compton-Y is significantly more skewed than what was found for the stellar and gas mass by \citet{Farahi2018}. However, for Compton-Y the deviations from a lognormal depend on the size of the aperture, which in this work we take to be $5R_\text{500c}$ for Compton-Y as appropriate for the Planck satellite. In Appendix~\ref{sec:cyr500} we demonstrate that the deviations largely disappear when using a smaller aperture of $R_{\rm 500c}$, which suggests that the deviations visible in Fig.~\ref{fig:lognorm} are due to projection of/blending with nearby structures. 

For richness we do not attempt to fit a lognormal plus power-law, since this shape is not clearly seen in the distributions. For the highest mass bins the shape is lognormal, and for the lower mass bins there is a tail extending towards lower values of richness. 

For all three observables we find an increase in the lognormal scatter towards lower masses. We leave an investigation of the physical origin of the scatter in the different observables for future work.

\subsection{Correlations between cluster properties} \label{sec:general_correlations}
To better understand how different selections will relate to the different observables, we investigate the distributions of, and correlations between the observables we select on. In Fig.~\ref{fig:cornerplot} we show a corner plot of the distribution of our selection quantities at $z=0.3$. We pick an intermediate redshift, but note that the qualitative picture is similar at $z=0$ and $0.5$. The panels along the diagonal show histograms of the individual quantities. The off diagonal panels show the 95th percentiles for each combination of quantities. The light blue lines show the mass selected distribution for a lower limit on $M_{\rm 500c}$ of $M_\text{C} = 10^{14}~\rm{M}_{\odot}$. Each other colour shows the result for a different sample $A > A_\text{C}$, i.e.\ a selection based on the median value of the observable $A$ indicated in the legend at $A(M_{\rm 500c} = M_\text{C})$.

The light blue contours in the first column show that all observables are tightly correlated with halo mass for masses above the mass cut $M_{\rm 500c} = 10^{14}~\rm{M}_{\odot}$. For $M > M_\text{C}$ the differences between the different samples are small. Below this mass the distributions diverge. Richness shows the largest spread, and a cut of $M_{\rm 500c}=10^{14}~\rm{M}_{\odot}$ is still high enough for it to not suffer from small number statistics.

The mass distributions for selections based on the other quantities can be seen in the topmost diagonal panel. At the selection limit the number of objects selected based on the quantity shown along the $x$-axis of the histogram drops to zero. The richness selection includes the largest number of haloes below the target mass $M_\text{C}$ and starts to become incomplete, with respect to a mass selection, at masses below $\approx0.2$~dex above $M_\text{C}$. X-ray and Compton-Y selections are comparable to each other in terms of completeness at the target mass, and include less contamination from haloes with $M_{\rm 500c} < M_\text{C}$ than the sample selected on richness. At this redshift, X-ray selection yields the lowest number of haloes with mass smaller than the target mass.

\subsection{Characteristic mass as a function of the cut in observable space} \label{sec:characteristic_mass}
\begin{figure}
    \centering
    \includegraphics[width=.99\columnwidth]{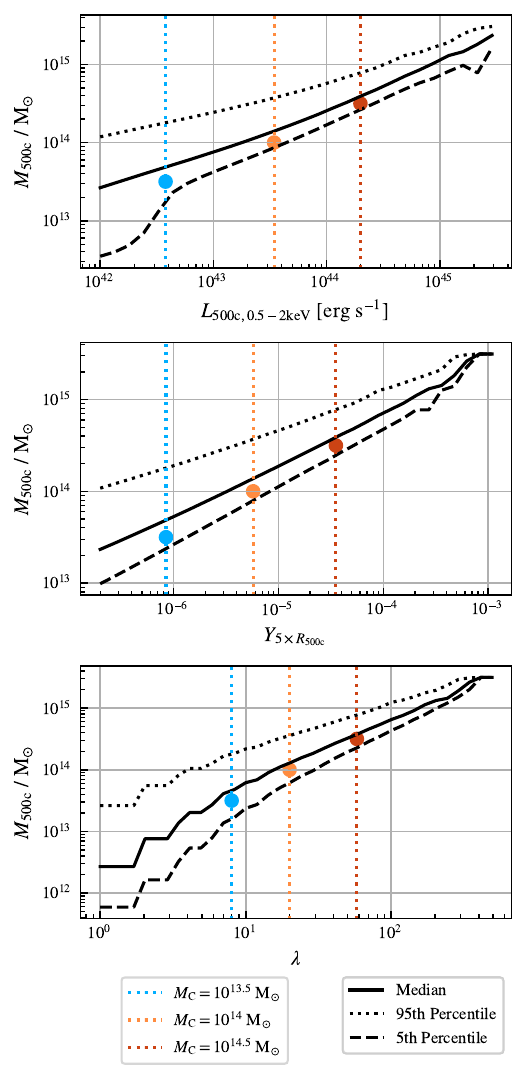}
    \caption{Solid lines show the median $M_{\rm 500c}$ for a sample selected using a cut on the quantity plotted along the $x$-axis, i.e.\ $\text{median}(M_{\rm 500c}|A>A_\text{C})$, for $A_{\rm C}$ given on the $x$-axis, at $z=0.3$. From top to bottom, the different panels show the three different selection quantities $A_\text{C}$: X-ray, Compton-Y and galaxy richness. The dashed (dotted) line indicates the 5th (95th) percentile. The vertical dotted lines show the median values of each quantity at the masses indicated in the legend, with the corresponding mass indicated with a dot. Except for richness at $\lambda < 10$, the median relations are smooth and have shapes close to power-laws.}
    \label{fig:fixedz}
\end{figure}

In addition to the complete distributions shown in Fig~\ref{fig:cornerplot}, it is interesting to look at each of the scaling relations between observable and mass that are used for the selection. The solid line in each panel of Fig.~\ref{fig:fixedz} shows the median $M_{500\rm c}$ for a sample defined by $A>A_\text{C}$ with $A_\text{C}$ plotted along the $x$-axis at $z=0.3$. Different panels show different choices for $A$. From top to bottom, the three panels show X-ray, SZ and richness selection. We also show the 5th and 95th percentiles of the sample, and for reference we indicate the median values at fixed mass, unlike the black lines which show the mass of the full sample with $A>A_\text{C}$, of the selection quantities at three fixed values of $M_\text{C}$ using vertical dotted lines, with a circle to mark the mass the line corresponds to.

In all panels, the median lines cross each vertical line at a mass that is slightly higher than the mass that the vertical line is based on, indicated by the circle. Since the vertical lines and circles indicate the value of $A$ for a sample with fixed mass $M_\text{C}$, while the black lines show the median based on the sample with $A> A(M_\text{C})$, this is expected. The difference is not very large, due to the exponential nature of the high-mass end of the halo mass function, every sample will be dominated by its lowest mass haloes. There is a very slight trend where for richness the crossing point is closest to the fixed mass $M_{\rm C}$ compared with X-ray and SZ selections. As seen in Fig,~\ref{fig:cornerplot}, richness starts becoming incomplete at a higher mass than the other selection methods, which will make the median mass in such a sample lower. 

Except for very low X-ray luminosity cuts, for all panels and all values of $A_\text{C}$ the median is closer to the 5th percentile than to the 95th percentile, indicating that the samples are skewed to lower halo masses. In Section~\ref{sec:lognorm} we showed that the intrinsic scatter in $A$ at fixed mass is largely consistent with an un-skewed lognormal distribution, with only slight deviations at the high-end tail. The skew we see in Fig.~\ref{fig:fixedz} is due to the nature of the selection. Because there are more lower-mass haloes with relatively high values of $A$ for their mass than there are higher-mass haloes with relatively low values of $A$ for their mass, up scatter dominates over down scatter. 

For most of the dynamic range shown in Fig.~\ref{fig:fixedz}, all percentiles have a smooth, near power-law shape, with two exceptions. First, at low X-ray luminosities there is a sudden drop in the 5th percentile, indicating a large amount of scatter of the X-ray luminosity in haloes with masses $M_\text{500c} < 10^{13.5}~\rm{M}_{\odot}$. When we do not mask particles recently heated by AGN, the drop of the percentile moves to a higher X-ray luminosity. This suggests that for low halo masses increases in X-ray luminosity due to feedback are important. In Appendix~\ref{sec:convergence} we show that the drop in the 5th percentile does not disappear for a simulation with higher resolution, and is thus not a resolution effect. From Fig.~\ref{fig:lognorm} we know that for X-ray luminosity the importance of up-scatter increases for lower halo masses, as the distribution at fixed mass gains a tail towards higher X-ray luminosities. In particular, this deviation from lognormal is larger for X-ray than for SZ. Our findings in Fig.~\ref{fig:fixedz} indicate that the X-ray deviations from lognormal are strong enough to significantly skew the sample at masses $M_{\rm 500c}<10^{13.5}~\rm{M}_{\odot}$. Second, for richness there is a clear deviation from the power-law shape for $\lambda < 10$. In addition, discreteness effects appear because a halo mass of about $10^{13}~\rm{M}_{\odot}$ is required for the richness to be larger than one. This behaviour is not affect by the resolution of the simulation, see Appendix~\ref{sec:convergence}, but does move to lower masses for higher resolutions.

\begin{figure*}
    \centering
    \includegraphics[width=\textwidth]{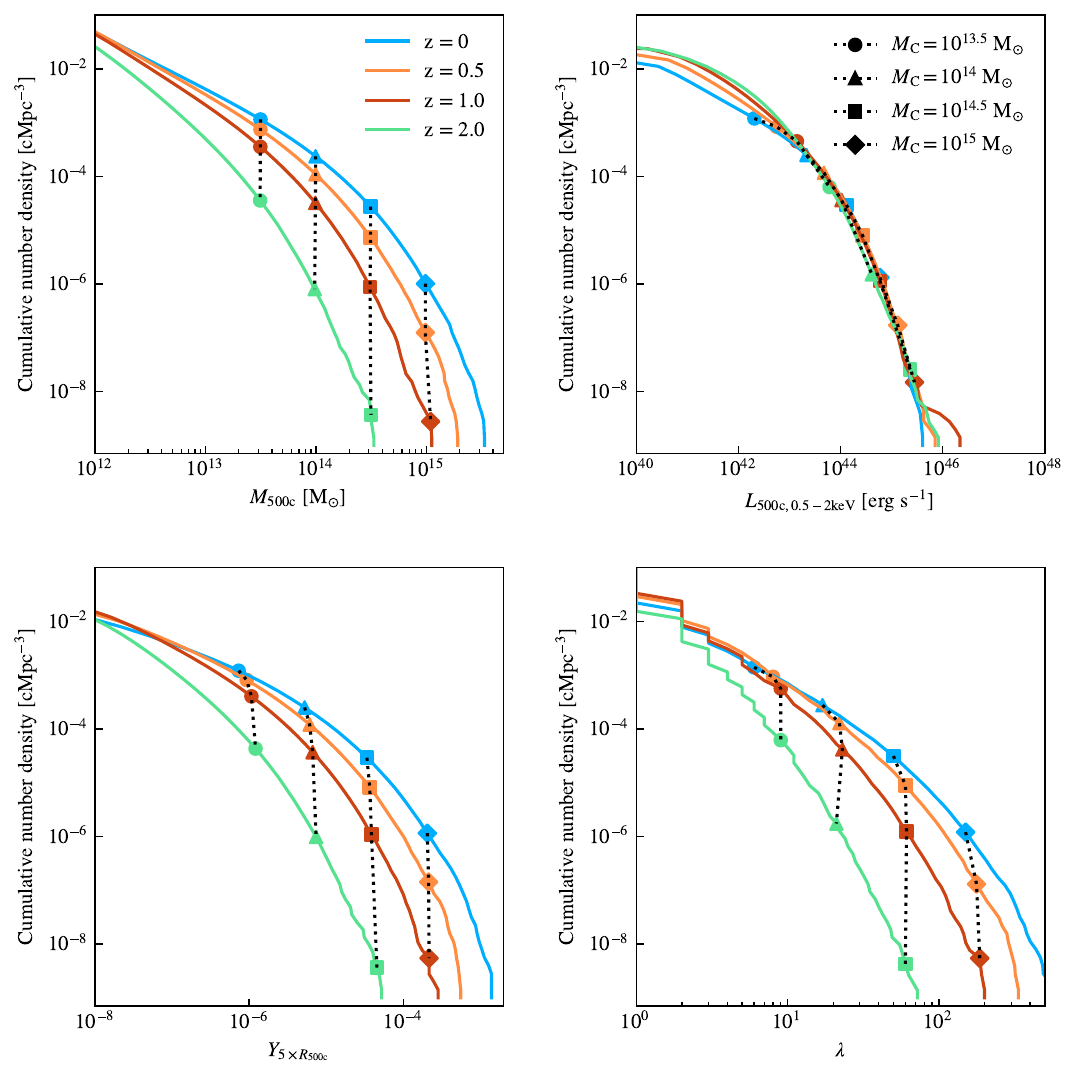}
    \caption{The cumulative comoving number density of all haloes as a function of $M_{\rm 500c}$ (top left panel), X-ray luminosity (top right panel), SZ Compton-Y (bottom left panel) and galaxy richness (bottom right panel). The different colours correspond to different redshifts. The symbols indicate the number density and median value of each quantity at a fixed mass, with the different symbols corresponding to different masses. At higher redshifts, a cut at fixed comoving number density results in a sample with lower masses, Compton-Y and richness, but a distribution of X-ray luminosities that is almost independent of redshift. The dotted lines connect the markers for the same mass at different redshifts. Because at a fixed mass the X-ray luminosity increases with redshift, the number density above a fixed luminosity decreases less with redshift than for a mass-selected sample. Compton-Y and richness selected samples evolve similarly to a selection based on mass.}
    \label{fig:numbden}
\end{figure*}

\subsection{Selection at fixed comoving number density} \label{sec:fixed_density}
In the previous subsection we created samples by making a cut on a selection observable, and then compared the resulting sample with a mass-selected sample. Another approach of interest is to create an ordered list based on the values of a selection quantity and then selecting a sample based on a cut in the cumulative comoving number density of objects. We show the cumulative comoving number density as a function of mass, X-ray luminosity, SZ signal and galaxy richness in the different panels of Fig.~\ref{fig:numbden}. Different colours correspond to different redshifts. Comparing the different coloured solid lines, we see that a cut on comoving cumulative number density corresponds, as redshift increases, to a sample with lower masses, Compton-Y values and richness values, but is close to an X-ray luminosity limited sample for number densities greater than $10^{-3}\,\text{cMpc}^{-3}$. 

Except for X-ray luminosity, the number density at a fixed value of the selection quantity decreases strongly with increasing redshift. For a mass-selected sample this is expected, because the halo mass function increases with time. For selection quantities for which the observable -- mass relation does not evolve strongly, we expect the same qualitative trend, which is indeed seen for the SZ signal and, to a lesser extent, galaxy richness. Interestingly, for X-ray luminosity the different redshifts fall nearly on top of each other, except at the faint end. This implies that the evolution of the luminosity -- mass relation nearly cancels the evolution of the mass function, with luminosity at fixed mass increasing with redshift. The very close agreement between the different redshifts must be a coincidence, because the number density -- mass relation depends differently and more strongly on cosmology than the observable - mass relation. Note that, as a consequence, to create a mass-selected sample, we would need to select much higher X-ray luminosities, slightly higher Compton-Y values, and much smaller richness values at high redshift compared to $z=0$. 

It is helpful to consider the symbols connected with dotted lines, which inform us about the evolution of the selection quantity at fixed $M_\text{500c}$. For the SZ Compton-Y the dotted curves are nearly vertical, which implies that there is very little evolution in the mass -- observable relation. For SZ the curves have negative slopes, bending slightly towards lower values at higher number densities. Because number density increases with time at fixed mass, this indicates a slight evolution towards smaller Compton-Y at fixed mass, as expected from the $E^{3/2}(z)$ scaling from self similarity \citep{Kaiser1986,Kaiser1991}. For X-ray luminosity the dotted lines bend strongly in the same direction, implying strong evolution towards lower luminosities at fixed mass, as expected from the $E^{2}(z)$ self-similar scaling. For galaxy richness the dotted curves behave similarly to Compton-Y, but slow slightly more evolution with redshift.

\subsection{Sample mass bias as a function of the selection limit} \label{sec:fixed_mass}
\begin{figure}
    \centering
    \includegraphics[width=.93\columnwidth]{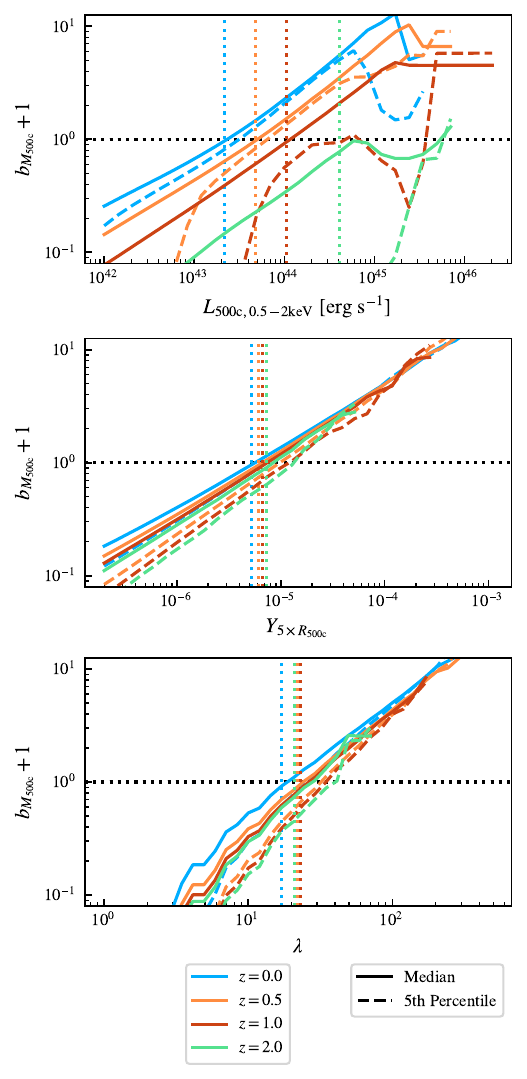}
    \caption{Each panel shows the bias factor for $M_{\rm 500c}$ (Eq.~\ref{eq:samplebias}) of a different sample relative to a mass-selected sample with a mass cut of $M_\text{C}=10^{14}~\text{M}_{\odot}$. We add one to the sample mass bias to allow for logarithmic plotting. The panels show samples selected to have $A>a$, and thus contain all halos above the given threshold $a$, where $A$ is X-ray luminosity (top panel), SZ Compton-Y (middle panel), or galaxy richness (bottom panel), and $a$ is the value plotted along the $x$-axis. Solid and dashed lines show the bias for the median and the 5th percentile of the distribution, respectively. The different colours show the results for four different redshifts. The dotted vertical lines show the median value of each quantity at the mass $M_\text{C}$. For an unbiased sample both the solid and dashed curves would intersect the vertical dotted line at the $y$-axis value of unity. }
    \label{fig:fixedmass}
\end{figure}

The next step is to see how the sample mass bias changes with the selection limit $A>a$ and how it evolves with redshift. To indicate how different the sample is from a mass-selected sample with a mass cut $M_\text{C}$, which we will hold fixed at $10^{14}~\rm{M}_{\odot}$, we compute the sample mass bias, as defined by Eq.~\ref{eq:samplebias}, where we add one to the bias to allow for logarithmic plotting. The solid lines in Fig.~\ref{fig:fixedmass} show the sample mass bias for the median, i.e.\ the factor by which the median mass of the sample with $A>a$, where $a$ is plotted along the $x$-axis, differs from the median mass of the sample with $M_\text{500c} > M_\text{C}$. Similarly, the dashed lines show the sample mass bias for the 5th percentile. The different colours show  different redshifts. The three panels show selections based on X-ray luminosity (top), SZ signal (middle) and galaxy richness (bottom). The bias is defined with the respect to the sample with a mass cut of $M_{\rm C}=10^{14}~\rm{M}_{\odot}$. 

For reference, the median values of observable $A$ at the fixed mass $M_\text{C}$ are indicated by the dotted vertical lines, one for each redshift. The vertical lines show strong redshift evolution of the value of the median X-ray luminosity at mass $M_\text{C}$, with the median luminosity increasing by over an order of magnitude from $z=0$ to $z=2$. For the SZ signal the effect is much milder, there is only a slight increase with redshift. Galaxy richness only exhibits mild evolution.

Observed clusters are distributed across a range of redshifts. If the observable-mass relation evolves, then applying a cut at a single value of the observable $a$ can result in samples for which the mass distribution varies with redshift. This then leads to different sample mass biases for different redshifts. This effect is most pronounced for X-ray selection, as can be seen from the large differences between the different coloured solid lines. For example, while choosing a luminosity cut of $2\times 10^{43}~\text{erg}\,\text{s}^{-1}$ yields a sample with a nearly unbiased median mass at $z=0$, while at $z=2$ the median mass is biased low by nearly an order of magnitude. Due to the strong evolution in the relation between X-ray luminosity and mass, any value selected for the X-ray luminosity cut will lead to a sample that becomes increasingly biased towards lower masses at higher redshifts. On the other hand, thanks to the mild redshift evolution for the SZ signal and galaxy richness, a cut on Compton-Y or $\lambda$ will lead to a similar mass cut across different redshifts, thus allowing for the creation of a relatively unbiased sample. For a fixed cut in the observable, the value of the sample mass bias decreases with redshift for X-ray luminosity and SZ signal, but tends to increase with redshift for galaxy richness. 

Examining the 5th percentiles, we see that they yield lower sample mass bias factors than for the medians (i.e.\ the dashed lines are below the solid lines of the same colour), indicating the sample is skewed towards lower masses. For cuts resulting in an unbiased median (i.e.\ samples with $A>a$ where the value $a$ corresponds to the intersect of the vertical coloured dotted line and the horizontal black dotted line indicating $b_{\rm M_{500c}}=1$), the 5th percentile is biased low (i.e.\ the dashed line of the corresponding colour gives a bias value lower than unity). This means that the 5th percentile of the mass distribution of the sample with $A>a$, where $a$ is chosen such that the median mass is the same as for a sample with $M>M_\text{C}$, is smaller than the 5th percentile of this mass-selected sample. This bias tends to increase with redshift and becomes particularly large for X-ray selection at $z=2$. 

Below a certain X-ray luminosity, the bias factor for the 5th percentile decreases rapidly to values $b_{\rm M_{500c}}\ll 10^{-1}$. This suggests that for low halo masses, there is a large amount of scatter in the X-ray luminosity. This behaviour is similar to that for X-ray selection shown in Fig.~\ref{fig:fixedz}. The sudden drop in the bias shifts to higher luminosities at higher redshifts. There are no similar drops in the bias factor for SZ or richness selection. 

\subsection{Sample mass bias as a function of the target mass limit} \label{sec:changing_mass}

\begin{figure}
    \centering
    \includegraphics[width=\columnwidth]{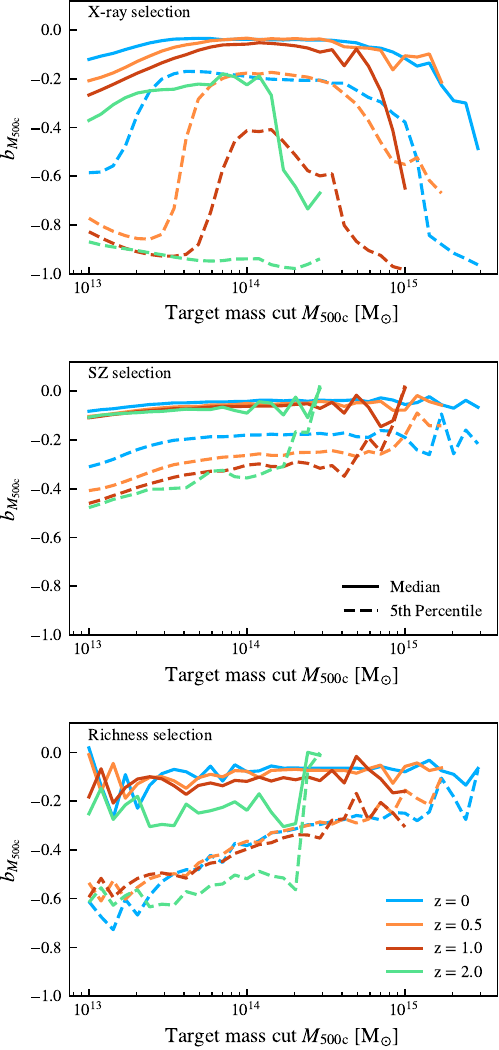}
    \caption{The sample $M_{500\rm{c}}$ bias (Eq.~\ref{eq:samplebias}) as a function of the target mass cut. The cut used for X-ray luminosity (top panel), Compton-Y (middle panel) and galaxy richness (bottom panel) is the median value for the mass cut plotted along the $x$-axis (see Eq.~\ref{eq:cut_eq}). The solid and dashed lines shows the sample mass bias for the median and 5th percentile, respectively. The different line colours show the results at different redshifts. The bias in the median mass increases towards low target masses and, for X-ray selection also towards high masses. While there are target masses for which the median mass is only slightly biased low, the 5th percentiles of the mass distribution are always much lower than for a mass-selected sample. The biases tend to increase with redshift.}
    \label{fig:changing mass}
\end{figure}

Next we will investigate the bias in the median and 5th percentile $M_{\rm 500c}$ for samples created with different observables as a function of the target mass $M_{\rm C}$. To calculate the bias we use a cut based on the median of observable $a$ at a fixed mass $M_{\rm C}$ (Eq.~\ref{eq:cut_eq}), which we denote as $A_{\rm C}$. We then calculate the sample mass bias $b_{M_{\rm 500c}}$ for a range of $M_{\rm C}$ using Eq.~\ref{eq:samplebias}. Since we use $M_{\rm C}$ to define the cut $A_{\rm C}$, the sample mass bias becomes a function of only the target mass. This is shown for different observables in Fig.~\ref{fig:changing mass}. From top to bottom, the mass cut is informed by a X-ray, SZ or richness selection limit. Each panel uses four distinct colours to represent various redshifts. The solid and dashed lines, respectively, depict the bias in the median and 5th percentile.

We first discuss some of the apparently odd features in each of the panels. Similar to what is shown in Figs.~\ref{fig:fixedz} and \ref{fig:fixedmass}, there is a drop in the 5th percentile for the X-ray selection at low masses. The mass at which this happens increases with redshift and is $~2\times10^{13}~\rm{M}_{\odot}$ at $z=0$, increasing by a factor five at $z=1$. Additionally, both biases exhibit a drop-off at the highest masses. This is caused by the fact that there are only very few halos for those mass bins. In that case even a few lower mass haloes that have a relatively high X-ray luminosity can quickly contaminate the sample and lead to a large bias. 

At low masses, slightly above $M_{\rm 500c}=10^{13}~\rm{M}_{\odot}$, the richness selection demonstrates a sawtooth-like behaviour. This behaviour is directly linked to the discreteness issues inherent in our definition of richness. Every discrete value for the richness will have a range of halo masses for which it is the median at fixed mass. For this range of mass the richness selected sample will not change, and all the change is due to the mass selection. For each value of the richness there is a mass cut value that maximises the bias, and moving away from this value will always lead to an increasing bias. The decrease is turned around when the richness cut goes to the next discrete value, and then it will suddenly start to increase. This inherently leads to the lines going up and down with sudden changes in slope, which is seen in the figure as a saw tooth.

Now we will discuss the behaviour of the bias for each of the selections, starting with X-ray. At all redshifts the bias in the median mass for X-ray selection has a similar shape. Around $\sim10^{14}~\rm{M}_{\odot}$, the median bias is closest to zero, indicating a relatively unbiased selection, and it remains mostly flat around that mass range. Towards the highest masses, the bias has a sudden drop. The bias also slowly moves away from one towards lower masses. For $z=1, 0.5$ and 0, the bias is close to -0.1 at the maximum, and only at $z=2$ does the best possible bias decrease to just below -0.2. The 5th percentile exhibits more extreme evolution, with the plateau of roughly constant bias diminishing with increasing mass. At $z=2$, the 5th percentile is consistently biased by a factor of ten or more. In X-ray luminosity-based selections, optimal results are thus achieved by choosing a cut that maintains the median halo mass above $\sim10^{14}~\rm{M}_{\odot}$. This not only minimises bias but also prevents significant skewness in the distribution, especially at the 5th percentile.

The SZ selection consistently yields a median bias close to zero for the median across all masses and redshifts. The median bias increases slightly towards $\sim10^{13}\rm{M}_{\odot}$ but stays above -0.2 for all redshifts. This is in agreement with the results from Fig.~\ref{fig:fixedmass}. The SZ selection has little evolution with mass, and consistently provides relatively unbiased results ($-0.1< b < 0$) for all redshifts. Significant evolution is observed for the 5th percentile, becoming more biased with increasing redshift. The bias in the 5th percentile shifts from $-0.25$ at $z=0$ to approximately $-0.5$ at $z=1$ and $2$. The 5th percentiles become increasingly biased when the mass cut falls below $\sim2\times10^{13}~\rm{M}_{\odot}$. These results are for our fiducial SZ aperture of $5R_\text{500c}$. In Appendix~\ref{sec:cyr500} we investigate the bias for a smaller aperture of $R_{\rm 500c}$. Using the smaller aperture the biases reduce further, leading to a nearly unbiased median over the entire mass range, and the bias in the 5th percentile reduces to only $\sim5$ per cent. An aperture of $R_\text{500c}$ thus leads to smaller biases for SZ selection.

With the exception of masses $M_{\rm 500c}\sim10^{13}~\rm{M}_{\odot}$ and at $z=2$, richness selection leads to a median bias close to -0.1 that decreases slightly up to $z=2$. At the lowest masses richness exhibits a slight sawtooth behaviour due to discreteness effects, but the bias does not drop significantly. At $z=2$ the bias drops slightly more, reaching a value of slightly less than -0.2. The most interesting behaviour is found in the bias of the 5th percentile. Over the entire mass range, the bias in the 5h percentile increases with mass, going from -0.6 for $M_{\rm 500c}=10^{13}~\rm{M}_{\odot}$, to around -0.2 for $M_{\rm 500c}=10^{15}~\rm{M}_{\odot}$. The 5th percentile also becomes more biased at $z=2$.

For masses below $10^{14}~\rm{M}_{\odot}$ as well as at $z\geq1$, using an SZ selection yields the least biased results for both the median and the 5th percentile. In those regimes, the X-ray selection exhibits a substantial influx of smaller haloes 'up-scattering' into the sample, resulting in a stronger bias. For richness the median sample mass bias is similar to the SZ selection, but there is a much larger skew in the 5th percentile. When we examine selections above $\sim10^{14}~\rm{M}_{\odot}$ at $z=0$ and $0.5$ the three selections exhibit closer bias values, and there is no longer a clear 'best' choice. Regarding the bias on the median, the only outlier occurs for masses close to and larger than $10^{15}~\rm{M}_{\odot}$ in the case of an X-ray selection. In this scenario, both the median and the 5th percentile exhibit significant bias, and opting for either an SZ or richness selection yields better results.

\begin{figure*}
    \centering
    \includegraphics[width=0.99\textwidth]{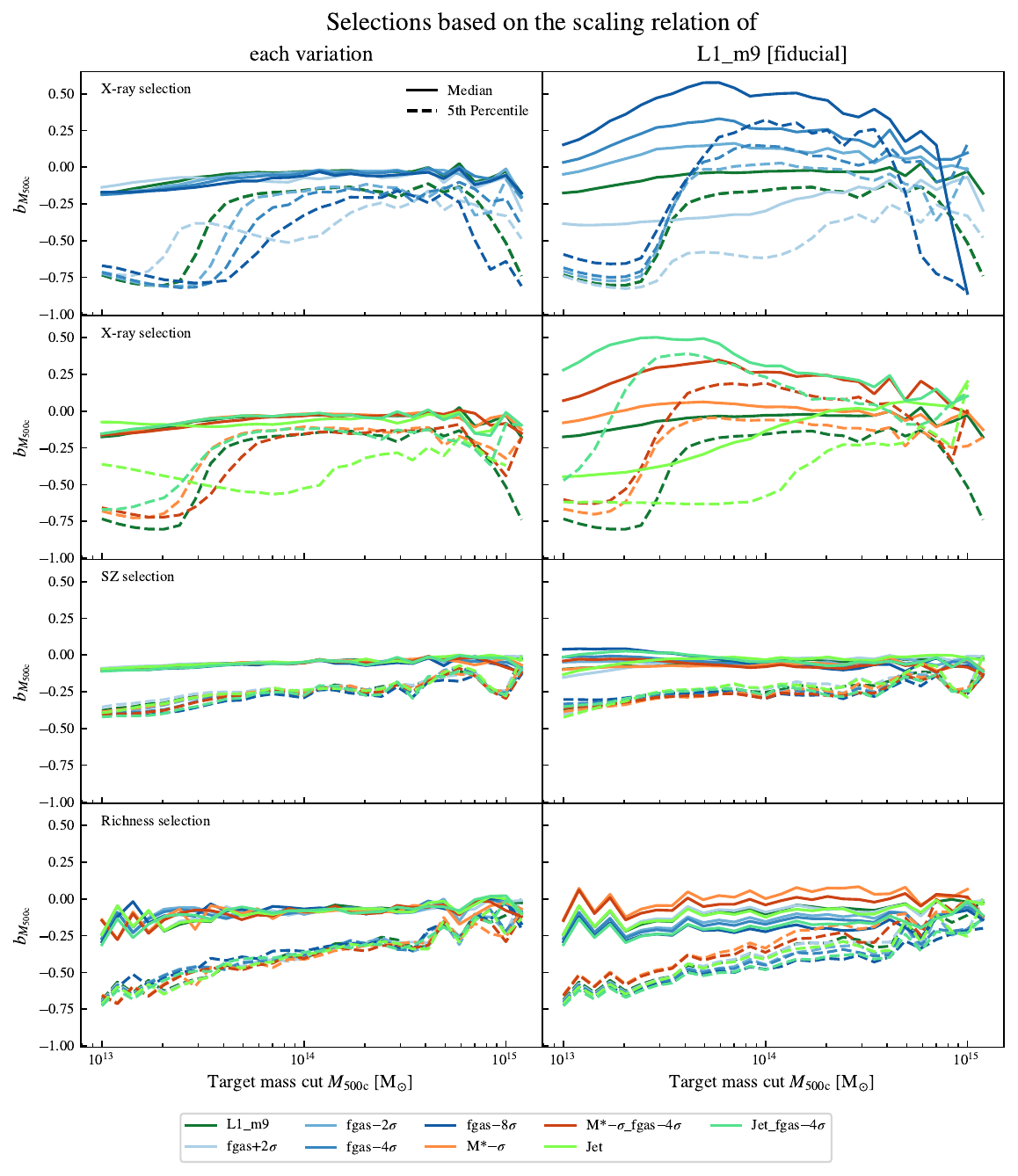}
    \caption{Same as Fig.~\ref{fig:changing mass} for a single redshift, $z=0.3$, but for different FLAMINGO models (different colours). In the left column the selection is based on the median of observable $a$ in each simulation at the target mass cut plotted along the horizontal axis. Differences between models are thus due to the differences in the scatter in the observable-mass relations. For all models shown in the right column the selection is based on the median relation of observable $a$ in the L1\_m9 simulation. Differences are therefore due to both changes in the scatter and changes in the median of the observable-mass relations. The labels ``fgas'' indicate runs with a change in the gas fractions, where a lower number of sigma indicates a lower gas fraction and thus stronger feedback. The M* label indicates runs for which the stellar mass function is shifted to lower masses. The label ``Jet'' indicates that the AGN feedback model uses kinetic jets instead of the thermally driven winds used for the other runs. Note that the top two rows both show results for X-ray selection, but for different sets of simulations. The results for X-ray selection are distributed over two panels for visual clarity. Only X-ray selection and, to a lesser extent, richness selection are sensitive to changes in the gas fraction or the stellar mass function.}
    \label{fig:feedbackvars}
\end{figure*}

\subsection{The effect of modelling uncertainty}
One potential reason for concern is that our conclusions might be influenced by the properties of clusters realised in the simulation and that these properties may not be modelled with sufficient accuracy. To examine the effect of varying the cluster properties, Fig.~\ref{fig:feedbackvars} shows the sample mass bias as a function of the target mass cut at $z=0.3$ for all the FLAMINGO feedback variations. As shown by \citep{FLAMINGOmain,Braspenning2023}, the cosmology variation have no significant impact on the scaling relations, and are therefore not considered in this work. These variations consist of models that vary the hot gas content and/or the stellar mass function, by changing the strengths of stellar and AGN feedback, or that use jet-like instead of thermal AGN feedback. In the left column, the X-ray luminosity, Compton-Y, and richness cuts correspond to the median value of the observable as a function of the target mass cut. For each model variation the cut therefore corresponds to the same target mass cut. In the right column we instead fix the X-ray, SZ and richness cuts to those obtained for the fiducial L1\_m9 simulation for the target mass cut. This translates to setting $A_{\rm C}(M_{\rm500c}) = A_{\rm C, L1\_m9}(M_{\rm 500c})$ for each variation, i.e.\ we assume a slightly wrong observable-mass scaling relation for the model variations. Therefore, the left column shows the effect of changes in the scatter in the observable-mass relation between the different models and the right columns shows the combined effect of changing the scatter and ignoring the effect of the change in model on the median observable-mass relation. 

Starting with the left column, which shows the effect of changing the scatter in the observable-mass relation, the results are similar for all model variations. Except for the 5th percentile of X-ray selected clusters for low target masses, the bias is generally insensitive to variations in the model. For X-ray there is a slight trend where a lower gas fraction (i.e.\ fgas$-N\sigma$) is associated with a slightly more biased median mass, but the effect is small. The shapes of the curves are different for the fgas+2$\sigma$ and Jet models, particularly for the bias on the 5th percentile.
For SZ- and richness-selected samples the bias factors are insensitive to the model. 

In the right column, which shows the combined effect of the model variation on the scatter and the mass-observable relation, we find larger though still small model-dependence for the sample mass bias in SZ-selected samples. The variations change the bias by $\Delta b_{M_\text{500c}} \approx 0.05-0.1$ and the general shape of the dependence on the target mass does not change. For richness-selection, we find a slight trend with gas fraction, and a deviation of about 0.1 in bias for the models with a stellar mass function shifted to lower stellar masses. The dependence on stellar mass is expected, as we apply a stellar mass cut for our definition or richness. In contrast with SZ, the differences between feedback variations are larger for higher mass objects. For X-ray selection, changing the simulation without changing the selection limit to account for the change in the mass-observable relation has a large impact. The bias on the median mass changes from $\approx 0.5$ to $\approx -0.5$ going from the lowest to highest fgas variation. This implies that having complete knowledge of the true scaling relation is essential. Any deviations between the true scaling relation and the one that is assumed when modelling selections effects will lead to a biased sample.

The fact that X-ray selection is most affected by variations in the model is to be expected. From \cite{Braspenning2023} we know that the different variations have different electron densities in the cluster cores. The X-ray luminosity scales as $\rho^2$ and is therefore more sensitive to feedback processes affecting the core than Compton-Y, which scales as $\rho$. From Fig.~7 of \citet{Kugel2023} or Fig.~10 of \citet{FLAMINGOmain} we can see that the gas fractions of all models start to converge for high masses, just as the sample mass bias start to converge for high masses in the top right panel of Figure~\ref{fig:feedbackvars}, though substantial differences remain even at the highest masses. It is also clear that the behaviour is not fully determined by the gas fraction, as the bias for the jet and the M* variations do not agree with their corresponding fgas variations. This further emphasises the fact that direct knowledge of the observable-mass scaling relations is important, and that we cannot rely solely on indirect measurements.

\begin{figure*}
    \centering
    \includegraphics[width=\textwidth]{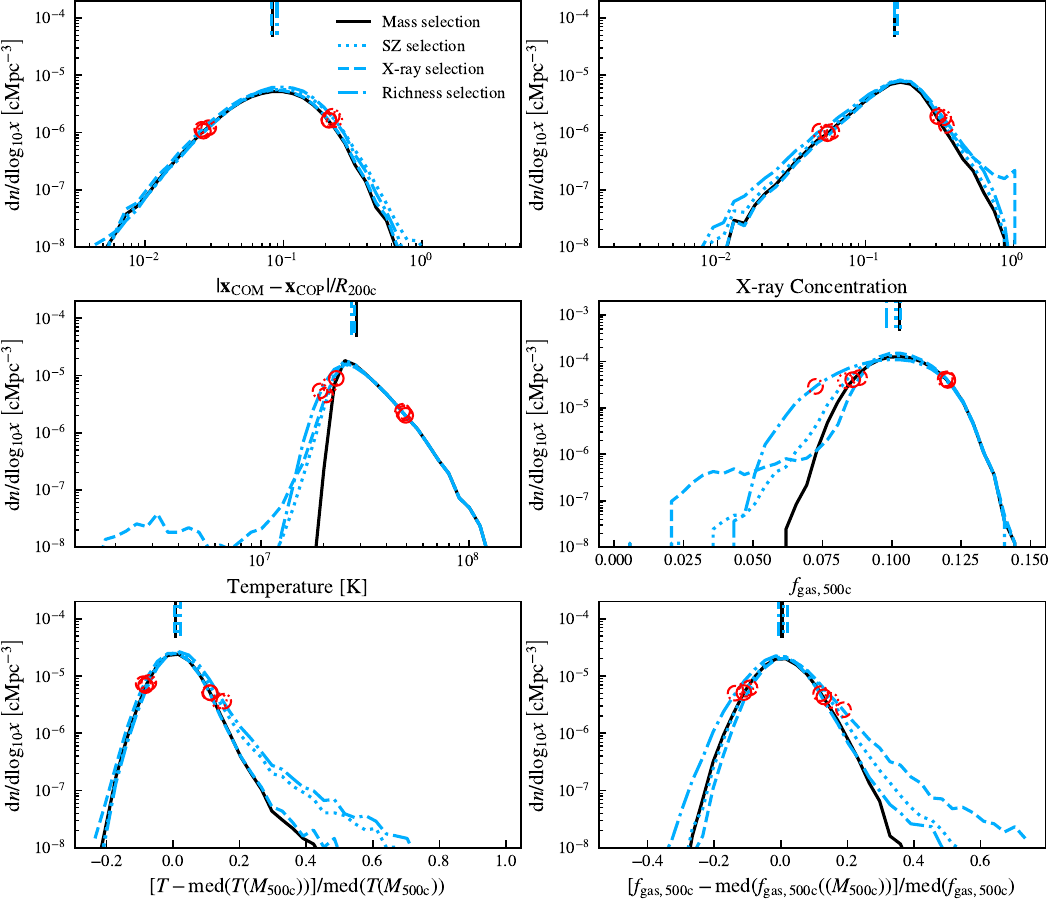}
    \caption{Distribution of relaxedness (top left; Eq.~\ref{eq:relaxedness}), X-ray concentration (top right; Eq.~\ref{eq:xconcentration}), mass-weighted mean temperature (middle left), gas fraction (middle right), relative deviation from the median mass-weighted mean temperature at fixed mass (bottom left) and relative deviation from the median $f_{\rm gas,500c}$ at fixed mass (bottom right) for selections with a target mass of $M_{\rm C}=10^{14}~\rm{M}_{\odot}$ at $z=0.3$. The different line styles indicate selections based on different properties. The median of each sample is indicated with a vertical line at the top of each panel. The 5th and 95th percentiles are shown using red circles. Selection effects result in biased distributions of cluster properties relative to a mass-selected sample. This is mostly due to upscatter from lower masses, but the bottom row shows that there are biases even at fixed true mass.}
    \label{fig:bias_quantities}
\end{figure*}

\subsection{Biases in properties other than mass}\label{sec:biased_props}
So far we have looked at how different selections bias the mass distribution of the cluster samples. When looking beyond the effects on cluster count cosmology, we want to inspect what the impact of different selections is on other properties of clusters. Even if the mass is measured independently, the lower mass objects that up-scatter into the selection could give a biased view of how scaling relations extrapolate towards lower masses. 

There are a few cluster properties that are of particular interest. \citet{Lovisari2017}, \citet{Rossetti2017} and \citet{Andrade2017} report differences in the disturbed fraction and the cool core fraction when comparing X-ray- and SZ-selected samples. Besides the disturbed fraction and cool core fraction, we also investigate biases in the median temperature and gas fraction. 

To quantify the degree of disturbedness in FLAMINGO, we compute the relaxedness parameter, defined as
\begin{equation} \label{eq:relaxedness}
    \text{Relaxedness}~=~\frac{|\mathbf{x}_{\rm{COM}}-\mathbf{x}_{\rm{COP}}|}{R_{\rm 200c}},
\end{equation}
where $\mathbf{x}_{\rm{COM}}$ is the position of the center of mass of the halo, defined by all the particles bound to the subhalo, $\mathbf{x}_{\rm{COP}}$ is the location of the most bound particle in the halo, and $R_{\rm 200c}$ is the radius within which the average density is equal to two hundred times the critical density. Note that a higher relaxedness value indicates a cluster that is more disturbed. 

In order to trace whether a cluster is cool-core, we use the X-ray concentration, defined as
\begin{equation} \label{eq:xconcentration}
    \text{X-ray concentration}~=~\frac{L_{X,r<0.15R_\text{500c}}}{L_{X,r<R_\text{500c}}},
\end{equation}
where $L_{X,r<0.15R_\text{500c}}$ is the X-ray luminosity in the core of the halo, defined by $0.15R_{\rm 500c}$ and $L_{X,r<R_{500}}$ is the total X-ray luminosity within $R_{\rm 500c}$. The higher the X-ray luminosity concentration, the more likely a cluster is to have a cool core.

We also measure the mass-weighted mean temperature, excluding gas below $10^{5}~\rm{K}$, and the gas mass fraction, each within $R_{\rm 500c}$. Additionally, since both the temperature and the gas fraction have a strong dependence on halo mass, we measure their deviations from the median at a fixed mass,
\begin{equation}\label{eq:med}
    \Delta X= \frac{X-\text{median}(X(M_{\rm 500c}))}{\text{median}(X(M_{\rm 500c}))}.
\end{equation}
This way we can investigate whether the lower mass haloes that up-scatter have different values for the temperature and gas fraction than a mass-selected sample.

To investigate how the different selections bias these quantities, we create a sample using a target mass $M_{\rm C}=10^{14}~\rm{M}_{\odot}$ for each observable $a$, as well as a mass-selected sample. In Figure~\ref{fig:bias_quantities} we show the distributions of these quantities at $z=0.3$. On the y-axis we show the bin-size normalised number density. The different line styles indicate the different selection methods used. The mass selection (black solid curve) should be taken as the baseline to compare the other selections with. We show the median of each selection with a vertical line at the top of the plot, and the 5th and 95th percentiles using red circles.

The top left panel shows the distribution of relaxedness, the offset between the center of potential and center of mass. We do not find strong differences between the different selection methods, the medians and percentiles are similar, and close to those of the mass selected sample. For the most disturbed objects, with the highest value of the offset, there is a slight trend where an SZ selection yield more highly disturbed objects, but this trend is very slight.

In the top right panel we show the distribution of X-ray concentrations. \citet{Andrade2017} used a similar metric to divide clusters into cool-core and non-cool-core clusters. A higher value indicates a more centrally concentrated X-ray luminosity, implying that the cluster is more likely to be a cool-core cluster. Richness selection does not lead to a clear preference between more or less cool-core objects. For X-ray and SZ we find results qualitatively similar to those of \citet{Andrade2017}. There is both an enhancement in the number of clusters with high X-ray concentration for the X-ray selection, and an enhancement of object with low X-ray concentration for SZ selection. However, as can been seen in the medians and percentiles, this difference is quite small.

The middle two panels of Figure~\ref{fig:bias_quantities} show the distributions of the temperature and gas fraction. For both these quantities, the differences relative to mass selection stem mainly from the fact that up-scattered haloes have lower halo masses, which implies that selections with more up-scattered haloes contain more objects with a low temperature and a low gas fraction. The most massive haloes, which have the highest temperatures and gas fractions, are included in each selection. This is reflected in the medians and 95th percentile, which do not change significantly, with the exception of the median gas fraction for richness selection. All selections are therefore complete for high temperatures and gas fractions. The samples selected on observables other than mass include more objects that have relatively low temperatures and gas fractions. For the temperature, the distribution of these objects is similar to what is found in Figure~\ref{fig:cornerplot}, indicating that the differences are largely mass-driven. These panels show that many of the lower-mass haloes that up-scatter into each selection have a significantly lower temperature and gas fraction than the haloes in a mass-selected sample. However, they do not tell us whether the haloes that are now included are different from other haloes of the same mass. This is investigated in the bottom two panels.

The bottom left panel of Figure~\ref{fig:bias_quantities} shows the relative deviation from the median temperature at the true halo mass for the different selections (see Eq.~\ref{eq:med}). By plotting this relative difference we can investigate how the temperatures are biased with respect to the median at their given mass. In this case the X-ray selection does not bias the sample substantially, but for the SZ and richness selections there is a pronounced tail towards haloes with much higher temperatures. This slightly increases the 95th percentile, but does not change the median significantly.

The bottom right panel of Figure~\ref{fig:bias_quantities} shows the deviation from the median gas fraction within $R_{\rm 500c}$ at fixed halo mass for the different selections. Richness selection increases the scatter, but there is no clear preference for higher or lower gas fractions as the median does not change. For both X-ray and SZ selections there is a preference for objects with  a gas fraction that is high for their mass, though the median gas fractions are nearly the same. X-ray selection finds slightly fewer haloes with a relatively low gas fraction than mass selection. This implies that even for haloes of a fixed mass, X-ray selection will already lead to a slight bias towards higher gas fractions. For both SZ and X-ray selection the clusters in the sample tend to have gas fractions that are higher than the average population, even at a fixed mass. For the X-ray sample, the 95th percentile increases by about $\sim20\%$ and higher percentiles are biased more strongly. This bias will be stronger closer to the survey selection limit, i.e.\ for lower masses. This is consistent with the findings by \citet{Kugel2023}, who attributed the fact that the observed relation between X-ray gas fraction and mass flattens off below $7\times10^{13}~\rm{M}_{\odot}$ to selection effects.

For three of the four quantities investigated, i.e.\ X-ray concentration, temperature, and gas fraction, we find that selecting on an observable other than mass introduces slight biases compared to a mass-selected sample. For clusters with masses larger than the median of the sample these effects will be negligible. However, upscatter results in the addition of lower mass haloes with temperatures and gas fractions that are lower than for a mass-selected sample. For richness selection this upscatter results in significant negative biases for the median gas fraction, while the bias in the medians is negligible for other selections. Even for the 5th percentiles the biases are small, with the exception of richness selection. 
Comparing the temperatures and gas fractions to the median values for the true mass of each selected halo, we find again that the medians are nearly unbiased, but there is a tail towards higher temperatures and gas fractions. For richness selection the up-scattered haloes also have a tail towards lower gas fractions relative to that expected for their true mass. However, for the 5th and 95th percentiles the biases are still small. As all our selections are intrinsically volume-limited, we find results similar to \citet{Chon2017}, and we note that observed differences between differently selected samples may be more influenced by the difference between volume- and flux-limited surveys than the selection method.

While these results cannot explain the relatively large sample mass biases that we found in earlier sections, they do show that some of the biases in cluster properties other than mass are intrinsically correlated with the chosen selection method.

\section{Conclusions}\label{sec:conc}
Given their large volumes and good agreement with observation, as well as the availability of a large number of model variations, the FLAMINGO simulations \citep{FLAMINGOmain,Kugel2023} provide an opportunity to investigate how different galaxy cluster selection methods influence the resulting samples. This is crucial for cluster cosmology \citep[e.g.][]{ClusterReviewAll2011,Mantz2019}, but also for understanding the role of selection biases in cluster scaling relations. 

We used the FLAMINGO simulations to investigate how the samples obtained from cuts in X-ray luminosity, thermal SZ Compton-Y (integrated within an aperture of $5R_\text{500c}$), or galaxy richness (using satellite galaxies with stellar mass $> 10^{10.046}\,\text{M}_\odot$) are biased in terms of the median and other percentiles of the mass distribution and certain secondary quantities. We summarise our findings as follows:
\begin{itemize}
    \item The scatter in X-ray luminosity, Compton-Y and richness increases with decreasing halo mass (see Fig.~\ref{fig:lognorm}). At fixed mass only the central parts of the distributions are lognormal. The distributions of X-ray luminosity and Compton-Y have power-law tails towards higher values, while for richness there can also be a tail towards lower values. The tails in the distributions cause the number of haloes that up-scatter into an X-ray or SZ selected sample to be underestimated when assuming lognormal scatter.
    \item In Fig.~\ref{fig:cornerplot} we compared the distributions of halo mass, X-ray luminosity, Compton-Y and richness for a target mass cut of $M_\text{500c} = 10^{14}~\rm{M}_{\odot}$ at $z=0.3$ for samples selected by mass or by $A>A_\text{C}$ where $A_\text{C} = \text{median}(A(M_{\rm 500c}=10^{14}~\rm{M}_{\odot}))$ and $A$ is the observable. We found tight correlations between all quantities for $A>A_\text{C}$, but not for lower values. Selecting based on richness leads to the largest amount of contamination by low-mass haloes, while X-ray selection yields the least amount of contamination. 
    \item As shown in Fig.~\ref{fig:fixedz}, increasing the selection limit in terms of X-ray luminosity, Compton-Y or richness leads to a sample with a smoothly increasing median and 95th percentile mass. However, for an X-ray luminosity cut smaller than $10^{43}~\rm{erg}\,\text{s}^{-1}$, the 5th percentile of the mass distribution dips to very low masses. This effect is converged with the numerical resolution (see Fig.~\ref{fig:convergence_plot}) and is qualitatively robust to changes in the subgrid feedback modelling (see Fig.~\ref{fig:feedbackvars}).
    \item The comoving number density above a fixed X-ray luminosity (richness) decreases less (more) with increasing redshift than for a mass-selected sample. A Compton-Y or richness selected sample evolves similarly to a selection based on mass (see Fig.~\ref{fig:numbden}).
    \item For a fixed target mass cut of $M_{\rm 500c}=10^{14}~\rm{M}_{\odot}$, the corresponding X-ray luminosity cut increases by more than an order of magnitude from $z=0$ to 2, while the richness cut decreases by about a factor of 3. For Compton-Y and richness the cut remains nearly constant with redshift (see Fig.~\ref{fig:fixedmass}).
    \item The bias in the median mass becomes stronger towards lower target masses and, for X-ray selection, also towards high masses. While there are target masses for which the median mass is only biased slightly low, the 5th percentiles of the mass distribution are always much lower than for a mass-selected sample. The samples tend to become more biased with increasing redshift. The target mass range for which the median sample mass bias is small is largest for SZ selection (see Fig.~\ref{fig:changing mass}). When using a $R_{\rm 500c}$ aperture instead of our fiducial $5R_{\rm 500c}$ aperture, the SZ selection becomes even less biased (Appendix~\ref{sec:cyr500}).
    \item Except for the 5th percentile of X-ray selected samples, and provided the median observable-mass relation is known, the bias factors are nearly the same for models calibrated to yield different gas fractions or stellar masses, and also for models using a different implementation of AGN feedback (Fig.~\ref{fig:feedbackvars}).
    \item The different selections lead to slight biases in cluster properties other than mass. In Figure~\ref{fig:bias_quantities} we demonstrated this for a target mass $M_{\rm 500c}=10^{14}~\rm{M}_{\odot}$. For X-ray selection, the lower mass objects that up-scatter into the sample have a very slight preference to have high X-ray concentrations, which is indicative of a cool core, while the opposite is true for selection based on richness. SZ selection includes slightly more clusters that are disturbed. Due to up-scatter of lower-mass haloes, all selections result in the inclusion of objects with temperatures and gas fractions that are much lower than are present in the mass-selected sample. However, compared with the median values for their true mass, the up-scattered objects tend to have high temperatures and gas fractions. Most of these effects are minor, leading to only small changes in the median and the 5th and 95th percentiles.
\end{itemize}

For each of the three selection methods, there are regimes in which the samples obtained have a small median sample mass bias. However, the 5th percentile of the mass distribution is nearly always biased significantly low and the biases tend to increase with redshift. Overall, SZ selection gives results that are closest to mass selection.

Overall, our results highlight how important it is that the scaling relations between mass and its observational proxies, including the scatter, are measured and modelled accurately. Even slight biases in the mass distributions can lead to differences that are problematic for surveys aimed at measuring cosmological parameters using cluster counts. We aim to investigate the direct effect of these biases on clusters counts in future work. 

We have shown that the objects with the lowest masses in each sample are more likely to be outliers with respect to the overall population when it comes to cluster properties other than the mass proxies. This can lead to biases when observationally determining scaling relations for quantities like the temperature and gas fractions.

In this work we have investigated selections based on observables in theory-space. We have ignored observational measurement errors, lightcone effects, projection effects, fore- and backgrounds, the effects of changing the cosmology, and other systematic effects, many of which will be survey specific. We have also implicitly assumed that observationally selection depends solely on the proxies investigated here, whereas in reality the signal-to-noise of a detection will depend on other properties. For example, X-ray selection likely depends not just on luminosity, but also on surface brightness \citep[see e.g.][]{andreon2024observed}. Galaxy richness does not rely on stellar mass selection, but depends on the luminosity and colour of the galaxies, as well as their distribution in phase space. It will be important to include such effects in future work, e.g.\ by forward modelling observational selection based on virtual observations created using the FLAMINGO lightcones. 

\section*{Acknowledgements}
This work is partly funded by research programme Athena 184.034.002 from the Dutch Research Council (NWO). This project has received funding from the European Research Council (ERC) under the European Union’s Horizon 2020 research and innovation programme (grant agreement No 769130). VJFM acknowledges support by NWO through the Dark Universe Science Collaboration (OCENW.XL21.XL21.025). This work used the DiRAC@Durham facility managed by the Institute for Computational Cosmology on behalf of the STFC DiRAC HPC Facility (www.dirac.ac.uk). The equipment was funded by BEIS capital funding via STFC capital grants ST/K00042X/1, ST/P002293/1, ST/R002371/1 and ST/S002502/1, Durham University and STFC operations grant ST/R000832/1. DiRAC is part of the National e-Infrastructure.

\section*{Data Availability}
All data presented in this paper will be shared upon reasonable request to the corresponding author.



\bibliographystyle{mnras}
\bibliography{example} 

\begin{thebibliography}{}
\makeatletter
\relax
\def\mn@urlcharsother{\let\do\@makeother \do\$\do\&\do\#\do\^\do\_\do\%\do\~}
\def\mn@doi{\begingroup\mn@urlcharsother \@ifnextchar [ {\mn@doi@}
  {\mn@doi@[]}}
\def\mn@doi@[#1]#2{\def\@tempa{#1}\ifx\@tempa\@empty \href
  {http://dx.doi.org/#2} {doi:#2}\else \href {http://dx.doi.org/#2} {#1}\fi
  \endgroup}
\def\mn@eprint#1#2{\mn@eprint@#1:#2::\@nil}
\def\mn@eprint@arXiv#1{\href {http://arxiv.org/abs/#1} {{\tt arXiv:#1}}}
\def\mn@eprint@dblp#1{\href {http://dblp.uni-trier.de/rec/bibtex/#1.xml}
  {dblp:#1}}
\def\mn@eprint@#1:#2:#3:#4\@nil{\def\@tempa {#1}\def\@tempb {#2}\def\@tempc
  {#3}\ifx \@tempc \@empty \let \@tempc \@tempb \let \@tempb \@tempa \fi \ifx
  \@tempb \@empty \def\@tempb {arXiv}\fi \@ifundefined
  {mn@eprint@\@tempb}{\@tempb:\@tempc}{\expandafter \expandafter \csname
  mn@eprint@\@tempb\endcsname \expandafter{\@tempc}}}

\bibitem[\protect\citeauthoryear{{Abbott} et~al.,}{{Abbott}
  et~al.}{2022}]{DESYR3}
{Abbott} T.~M.~C.,  et~al., 2022, \mn@doi [\prd] {10.1103/PhysRevD.105.023520},
  \href {https://ui.adsabs.harvard.edu/abs/2022PhRvD.105b3520A} {105, 023520}

\bibitem[\protect\citeauthoryear{{Allen}, {Evrard}  \& {Mantz}}{{Allen}
  et~al.}{2011}]{ClusterReviewAll2011}
{Allen} S.~W.,  {Evrard} A.~E.,   {Mantz} A.~B.,  2011, \mn@doi [\araa]
  {10.1146/annurev-astro-081710-102514}, \href
  {https://ui.adsabs.harvard.edu/abs/2011ARA&A..49..409A} {49, 409}

\bibitem[\protect\citeauthoryear{{Andrade-Santos} et~al.,}{{Andrade-Santos}
  et~al.}{2017}]{Andrade2017}
{Andrade-Santos} F.,  et~al., 2017, \mn@doi [\apj] {10.3847/1538-4357/aa7461},
  \href {https://ui.adsabs.harvard.edu/abs/2017ApJ...843...76A} {843, 76}

\bibitem[\protect\citeauthoryear{Andreon, Trinchieri  \& Moretti}{Andreon
  et~al.}{2024}]{andreon2024observed}
Andreon S.,  Trinchieri G.,   Moretti A.,  2024, Observed abundance of X-ray
  low surface brightness clusters in optical, X-ray, and SZ selected samples
  (\mn@eprint {arXiv} {2404.12435})

\bibitem[\protect\citeauthoryear{{Angulo}, {Springel}, {White}, {Jenkins},
  {Baugh}  \& {Frenk}}{{Angulo} et~al.}{2012}]{Angulo2012}
{Angulo} R.~E.,  {Springel} V.,  {White} S.~D.~M.,  {Jenkins} A.,  {Baugh}
  C.~M.,   {Frenk} C.~S.,  2012, \mn@doi [\mnras]
  {10.1111/j.1365-2966.2012.21830.x}, \href
  {https://ui.adsabs.harvard.edu/abs/2012MNRAS.426.2046A} {426, 2046}

\bibitem[\protect\citeauthoryear{{Aric{\`o}}, {Angulo}, {Contreras},
  {Ondaro-Mallea}, {Pellejero-Iba{\~n}ez}  \& {Zennaro}}{{Aric{\`o}}
  et~al.}{2021}]{Baryonifsim2021}
{Aric{\`o}} G.,  {Angulo} R.~E.,  {Contreras} S.,  {Ondaro-Mallea} L.,
  {Pellejero-Iba{\~n}ez} M.,   {Zennaro} M.,  2021, \mn@doi [\mnras]
  {10.1093/mnras/stab1911}, \href
  {https://ui.adsabs.harvard.edu/abs/2021MNRAS.506.4070A} {506, 4070}

\bibitem[\protect\citeauthoryear{{Artis}, {Melin}, {Bartlett}, {Murray}  \&
  {Euclid Consortium}}{{Artis} et~al.}{2022}]{Euclidclusters2022}
{Artis} E.,  {Melin} J.~B.,  {Bartlett} J.~G.,  {Murray} C.,   {Euclid
  Consortium} 2022, in mm Universe @ NIKA2 - Observing the mm Universe with the
  NIKA2 Camera. p. 00004 (\mn@eprint {arXiv} {2111.05432}),
  \mn@doi{10.1051/epjconf/202225700004}

\bibitem[\protect\citeauthoryear{{Bah{\'e}} et~al.,}{{Bah{\'e}}
  et~al.}{2017}]{Hydranga2017}
{Bah{\'e}} Y.~M.,  et~al., 2017, \mn@doi [\mnras] {10.1093/mnras/stx1403},
  \href {https://ui.adsabs.harvard.edu/abs/2017MNRAS.470.4186B} {470, 4186}

\bibitem[\protect\citeauthoryear{{Bah{\'e}} et~al.,}{{Bah{\'e}}
  et~al.}{2022}]{bahe2021}
{Bah{\'e}} Y.~M.,  et~al., 2022, \mn@doi [\mnras] {10.1093/mnras/stac1339},
  \href {https://ui.adsabs.harvard.edu/abs/2022MNRAS.516..167B} {516, 167}

\bibitem[\protect\citeauthoryear{{Barnes}, {Kay}, {Henson}, {McCarthy},
  {Schaye}  \& {Jenkins}}{{Barnes} et~al.}{2017}]{MACSIS2017}
{Barnes} D.~J.,  {Kay} S.~T.,  {Henson} M.~A.,  {McCarthy} I.~G.,  {Schaye} J.,
    {Jenkins} A.,  2017, \mn@doi [\mnras] {10.1093/mnras/stw2722}, \href
  {https://ui.adsabs.harvard.edu/abs/2017MNRAS.465..213B} {465, 213}

\bibitem[\protect\citeauthoryear{{Black} \& {Evrard}}{{Black} \&
  {Evrard}}{2022}]{Reddragon2022}
{Black} W.~K.,  {Evrard} A.,  2022, \mn@doi [\mnras] {10.1093/mnras/stac2052},
  \href {https://ui.adsabs.harvard.edu/abs/2022MNRAS.516.1170B} {516, 1170}

\bibitem[\protect\citeauthoryear{{Bleem} et~al.,}{{Bleem}
  et~al.}{2024}]{Bleem2024}
{Bleem} L.~E.,  et~al., 2024, \mn@doi [The Open Journal of Astrophysics]
  {10.21105/astro.2311.07512}, \href
  {https://ui.adsabs.harvard.edu/abs/2024OJAp....7E..13B} {7, 13}

\bibitem[\protect\citeauthoryear{{Bocquet} et~al.,}{{Bocquet}
  et~al.}{2019}]{SPTclustercosmo2019}
{Bocquet} S.,  et~al., 2019, \mn@doi [\apj] {10.3847/1538-4357/ab1f10}, \href
  {https://ui.adsabs.harvard.edu/abs/2019ApJ...878...55B} {878, 55}

\bibitem[\protect\citeauthoryear{{Bocquet}, {Heitmann}, {Habib}, {Lawrence},
  {Uram}, {Frontiere}, {Pope}  \& {Finkel}}{{Bocquet}
  et~al.}{2020}]{MiraTitanHMF2020}
{Bocquet} S.,  {Heitmann} K.,  {Habib} S.,  {Lawrence} E.,  {Uram} T.,
  {Frontiere} N.,  {Pope} A.,   {Finkel} H.,  2020, \mn@doi [\apj]
  {10.3847/1538-4357/abac5c}, \href
  {https://ui.adsabs.harvard.edu/abs/2020ApJ...901....5B} {901, 5}

\bibitem[\protect\citeauthoryear{{Bocquet} et~al.,}{{Bocquet}
  et~al.}{2024}]{Bocquet2024}
{Bocquet} S.,  et~al., 2024, \mn@doi [arXiv e-prints]
  {10.48550/arXiv.2401.02075}, \href
  {https://ui.adsabs.harvard.edu/abs/2024arXiv240102075B} {p. arXiv:2401.02075}

\bibitem[\protect\citeauthoryear{{Booth} \& {Schaye}}{{Booth} \&
  {Schaye}}{2009}]{BoothSchaye2009}
{Booth} C.~M.,  {Schaye} J.,  2009, \mn@doi [\mnras]
  {10.1111/j.1365-2966.2009.15043.x}, \href
  {https://ui.adsabs.harvard.edu/abs/2009MNRAS.398...53B} {398, 53}

\bibitem[\protect\citeauthoryear{{Borrow}, {Schaller}, {Bower}  \&
  {Schaye}}{{Borrow} et~al.}{2022}]{Sphenix2022}
{Borrow} J.,  {Schaller} M.,  {Bower} R.~G.,   {Schaye} J.,  2022, \mn@doi
  [\mnras] {10.1093/mnras/stab3166}, \href
  {https://ui.adsabs.harvard.edu/abs/2022MNRAS.511.2367B} {511, 2367}

\bibitem[\protect\citeauthoryear{{Braspenning} et~al.,}{{Braspenning}
  et~al.}{2023}]{Braspenning2023}
{Braspenning} J.,  et~al., 2023, \mn@doi [arXiv e-prints]
  {10.48550/arXiv.2312.08277}, \href
  {https://ui.adsabs.harvard.edu/abs/2023arXiv231208277B} {p. arXiv:2312.08277}

\bibitem[\protect\citeauthoryear{{Chaikin}, {Schaye}, {Schaller},
  {Ben{\'\i}tez-Llambay}, {Nobels}  \& {Ploeckinger}}{{Chaikin}
  et~al.}{2022a}]{Chaikin2022b}
{Chaikin} E.,  {Schaye} J.,  {Schaller} M.,  {Ben{\'\i}tez-Llambay} A.,
  {Nobels} F. S.~J.,   {Ploeckinger} S.,  2022a, \mn@doi [arXiv e-prints]
  {10.48550/arXiv.2211.04619}, \href
  {https://ui.adsabs.harvard.edu/abs/2022arXiv221104619C} {p. arXiv:2211.04619}

\bibitem[\protect\citeauthoryear{{Chaikin}, {Schaye}, {Schaller}, {Bah{\'e}},
  {Nobels}  \& {Ploeckinger}}{{Chaikin} et~al.}{2022b}]{Chaikin2022}
{Chaikin} E.,  {Schaye} J.,  {Schaller} M.,  {Bah{\'e}} Y.~M.,  {Nobels} F.
  S.~J.,   {Ploeckinger} S.,  2022b, \mn@doi [\mnras] {10.1093/mnras/stac1132},
  \href {https://ui.adsabs.harvard.edu/abs/2022MNRAS.514..249C} {514, 249}

\bibitem[\protect\citeauthoryear{{Chaubal} et~al.,}{{Chaubal}
  et~al.}{2022}]{Chaubal2022}
{Chaubal} P.~S.,  et~al., 2022, \mn@doi [\apj] {10.3847/1538-4357/ac6a55},
  \href {https://ui.adsabs.harvard.edu/abs/2022ApJ...931..139C} {931, 139}

\bibitem[\protect\citeauthoryear{{Chisari} et~al.,}{{Chisari}
  et~al.}{2019}]{Baryonification2019}
{Chisari} N.~E.,  et~al., 2019, \mn@doi [The Open Journal of Astrophysics]
  {10.21105/astro.1905.06082}, \href
  {https://ui.adsabs.harvard.edu/abs/2019OJAp....2E...4C} {2, 4}

\bibitem[\protect\citeauthoryear{{Chiu}, {Klein}, {Mohr}  \& {Bocquet}}{{Chiu}
  et~al.}{2023}]{eFEDSclustercosmo2023}
{Chiu} I.~N.,  {Klein} M.,  {Mohr} J.,   {Bocquet} S.,  2023, \mn@doi [\mnras]
  {10.1093/mnras/stad957}, \href
  {https://ui.adsabs.harvard.edu/abs/2023MNRAS.tmp..949C} {}

\bibitem[\protect\citeauthoryear{{Chon} \& {B{\"o}hringer}}{{Chon} \&
  {B{\"o}hringer}}{2017}]{Chon2017}
{Chon} G.,  {B{\"o}hringer} H.,  2017, \mn@doi [\aap]
  {10.1051/0004-6361/201731854}, \href
  {https://ui.adsabs.harvard.edu/abs/2017A&A...606L...4C} {606, L4}

\bibitem[\protect\citeauthoryear{{Clerc} et~al.,}{{Clerc}
  et~al.}{2024}]{Clerc2024}
{Clerc} N.,  et~al., 2024, \mn@doi [arXiv e-prints]
  {10.48550/arXiv.2402.08457}, \href
  {https://ui.adsabs.harvard.edu/abs/2024arXiv240208457C} {p. arXiv:2402.08457}

\bibitem[\protect\citeauthoryear{{Costanzi} et~al.,}{{Costanzi}
  et~al.}{2019}]{SDSSClustercosmo2019}
{Costanzi} M.,  et~al., 2019, \mn@doi [\mnras] {10.1093/mnras/stz1949}, \href
  {https://ui.adsabs.harvard.edu/abs/2019MNRAS.488.4779C} {488, 4779}

\bibitem[\protect\citeauthoryear{{Cui} et~al.,}{{Cui}
  et~al.}{2018}]{Threehundred2018}
{Cui} W.,  et~al., 2018, \mn@doi [\mnras] {10.1093/mnras/sty2111}, \href
  {https://ui.adsabs.harvard.edu/abs/2018MNRAS.480.2898C} {480, 2898}

\bibitem[\protect\citeauthoryear{{Dalla Vecchia} \& {Schaye}}{{Dalla Vecchia}
  \& {Schaye}}{2008}]{DVSchaye2008kin}
{Dalla Vecchia} C.,  {Schaye} J.,  2008, \mn@doi [\mnras]
  {10.1111/j.1365-2966.2008.13322.x}, \href
  {https://ui.adsabs.harvard.edu/abs/2008MNRAS.387.1431D} {387, 1431}

\bibitem[\protect\citeauthoryear{{Dav{\'e}}, {Angl{\'e}s-Alc{\'a}zar},
  {Narayanan}, {Li}, {Rafieferantsoa}  \& {Appleby}}{{Dav{\'e}}
  et~al.}{2019}]{Simba2019}
{Dav{\'e}} R.,  {Angl{\'e}s-Alc{\'a}zar} D.,  {Narayanan} D.,  {Li} Q.,
  {Rafieferantsoa} M.~H.,   {Appleby} S.,  2019, \mn@doi [\mnras]
  {10.1093/mnras/stz937}, \href
  {https://ui.adsabs.harvard.edu/abs/2019MNRAS.486.2827D} {486, 2827}

\bibitem[\protect\citeauthoryear{{Debackere}, {Schaye}  \&
  {Hoekstra}}{{Debackere} et~al.}{2021}]{Stijn2021}
{Debackere} S. N.~B.,  {Schaye} J.,   {Hoekstra} H.,  2021, \mn@doi [\mnras]
  {10.1093/mnras/stab1326}, \href
  {https://ui.adsabs.harvard.edu/abs/2021MNRAS.505..593D} {505, 593}

\bibitem[\protect\citeauthoryear{{Debackere}, {Hoekstra}, {Schaye}, {Heitmann}
  \& {Habib}}{{Debackere} et~al.}{2022a}]{Stijn2022a}
{Debackere} S. N.~B.,  {Hoekstra} H.,  {Schaye} J.,  {Heitmann} K.,   {Habib}
  S.,  2022a, \mn@doi [\mnras] {10.1093/mnras/stac1687}, \href
  {https://ui.adsabs.harvard.edu/abs/2022MNRAS.515.3383D} {515, 3383}

\bibitem[\protect\citeauthoryear{{Debackere}, {Hoekstra}  \&
  {Schaye}}{{Debackere} et~al.}{2022b}]{Stijn2022b}
{Debackere} S. N.~B.,  {Hoekstra} H.,   {Schaye} J.,  2022b, \mn@doi [\mnras]
  {10.1093/mnras/stac2077}, \href
  {https://ui.adsabs.harvard.edu/abs/2022MNRAS.515.6023D} {515, 6023}

\bibitem[\protect\citeauthoryear{{Driver} et~al.,}{{Driver}
  et~al.}{2022}]{Driver2022}
{Driver} S.~P.,  et~al., 2022, \mn@doi [\mnras] {10.1093/mnras/stac472}, \href
  {https://ui.adsabs.harvard.edu/abs/2022MNRAS.513..439D} {513, 439}

\bibitem[\protect\citeauthoryear{{Elbers}, {Frenk}, {Jenkins}, {Li}  \&
  {Pascoli}}{{Elbers} et~al.}{2021}]{Deltaf2021}
{Elbers} W.,  {Frenk} C.~S.,  {Jenkins} A.,  {Li} B.,   {Pascoli} S.,  2021,
  \mn@doi [\mnras] {10.1093/mnras/stab2260}, \href
  {https://ui.adsabs.harvard.edu/abs/2021MNRAS.507.2614E} {507, 2614}

\bibitem[\protect\citeauthoryear{{Elbers}, {Frenk}, {Jenkins}, {Li}  \&
  {Pascoli}}{{Elbers} et~al.}{2022}]{ElbersIcs2022}
{Elbers} W.,  {Frenk} C.~S.,  {Jenkins} A.,  {Li} B.,   {Pascoli} S.,  2022,
  \mn@doi [\mnras] {10.1093/mnras/stac2365}, \href
  {https://ui.adsabs.harvard.edu/abs/2022MNRAS.516.3821E} {516, 3821}

\bibitem[\protect\citeauthoryear{{Evrard}, {Arnault}, {Huterer}  \&
  {Farahi}}{{Evrard} et~al.}{2014}]{Evrard2014}
{Evrard} A.~E.,  {Arnault} P.,  {Huterer} D.,   {Farahi} A.,  2014, \mn@doi
  [\mnras] {10.1093/mnras/stu784}, \href
  {https://ui.adsabs.harvard.edu/abs/2014MNRAS.441.3562E} {441, 3562}

\bibitem[\protect\citeauthoryear{{Farahi}, {Evrard}, {McCarthy}, {Barnes}  \&
  {Kay}}{{Farahi} et~al.}{2018}]{Farahi2018}
{Farahi} A.,  {Evrard} A.~E.,  {McCarthy} I.,  {Barnes} D.~J.,   {Kay} S.~T.,
  2018, \mn@doi [\mnras] {10.1093/mnras/sty1179}, \href
  {https://ui.adsabs.harvard.edu/abs/2018MNRAS.478.2618F} {478, 2618}

\bibitem[\protect\citeauthoryear{{Ferland} et~al.,}{{Ferland}
  et~al.}{2017}]{CLOUDY2017}
{Ferland} G.~J.,  et~al., 2017, \mn@doi [\rmxaa] {10.48550/arXiv.1705.10877},
  \href {https://ui.adsabs.harvard.edu/abs/2017RMxAA..53..385F} {53, 385}

\bibitem[\protect\citeauthoryear{{Ghirardini} et~al.,}{{Ghirardini}
  et~al.}{2024}]{Ghirardini2024}
{Ghirardini} V.,  et~al., 2024, \mn@doi [arXiv e-prints]
  {10.48550/arXiv.2402.08458}, \href
  {https://ui.adsabs.harvard.edu/abs/2024arXiv240208458G} {p. arXiv:2402.08458}

\bibitem[\protect\citeauthoryear{{Giles} et~al.,}{{Giles}
  et~al.}{2022}]{Giles2022}
{Giles} P.~A.,  et~al., 2022, \mn@doi [\mnras] {10.1093/mnras/stac2414}, \href
  {https://ui.adsabs.harvard.edu/abs/2022MNRAS.516.3878G} {516, 3878}

\bibitem[\protect\citeauthoryear{{Giri} \& {Schneider}}{{Giri} \&
  {Schneider}}{2021}]{Baryonification2_2021}
{Giri} S.~K.,  {Schneider} A.,  2021, arXiv e-prints, \href
  {https://ui.adsabs.harvard.edu/abs/2021arXiv210808863G} {p. arXiv:2108.08863}

\bibitem[\protect\citeauthoryear{{Grandis} et~al.,}{{Grandis}
  et~al.}{2020}]{Selec2020}
{Grandis} S.,  et~al., 2020, \mn@doi [\mnras] {10.1093/mnras/staa2333}, \href
  {https://ui.adsabs.harvard.edu/abs/2020MNRAS.498..771G} {498, 771}

\bibitem[\protect\citeauthoryear{{Grandis} et~al.,}{{Grandis}
  et~al.}{2021}]{Grandis2021}
{Grandis} S.,  et~al., 2021, \mn@doi [\mnras] {10.1093/mnras/stab869}, \href
  {https://ui.adsabs.harvard.edu/abs/2021MNRAS.504.1253G} {504, 1253}

\bibitem[\protect\citeauthoryear{{Hahn}, {Martizzi}, {Wu}, {Evrard}, {Teyssier}
   \& {Wechsler}}{{Hahn} et~al.}{2017}]{Hahn2017}
{Hahn} O.,  {Martizzi} D.,  {Wu} H.-Y.,  {Evrard} A.~E.,  {Teyssier} R.,
  {Wechsler} R.~H.,  2017, \mn@doi [\mnras] {10.1093/mnras/stx001}, \href
  {https://ui.adsabs.harvard.edu/abs/2017MNRAS.470..166H} {470, 166}

\bibitem[\protect\citeauthoryear{{Hahn}, {Rampf}  \& {Uhlemann}}{{Hahn}
  et~al.}{2021}]{Monofonic2021}
{Hahn} O.,  {Rampf} C.,   {Uhlemann} C.,  2021, \mn@doi [\mnras]
  {10.1093/mnras/staa3773}, \href
  {https://ui.adsabs.harvard.edu/abs/2021MNRAS.503..426H} {503, 426}

\bibitem[\protect\citeauthoryear{{Han}, {Cole}, {Frenk}, {Benitez-Llambay}  \&
  {Helly}}{{Han} et~al.}{2018}]{HBT2017}
{Han} J.,  {Cole} S.,  {Frenk} C.~S.,  {Benitez-Llambay} A.,   {Helly} J.,
  2018, \mn@doi [\mnras] {10.1093/mnras/stx2792}, \href
  {https://ui.adsabs.harvard.edu/abs/2018MNRAS.474..604H} {474, 604}

\bibitem[\protect\citeauthoryear{{Heymans} et~al.,}{{Heymans}
  et~al.}{2021}]{KIDS2021}
{Heymans} C.,  et~al., 2021, \mn@doi [\aap] {10.1051/0004-6361/202039063},
  \href {https://ui.adsabs.harvard.edu/abs/2021A&A...646A.140H} {646, A140}

\bibitem[\protect\citeauthoryear{{Hirschmann}, {Dolag}, {Saro}, {Bachmann},
  {Borgani}  \& {Burkert}}{{Hirschmann} et~al.}{2014}]{Magneticum2014}
{Hirschmann} M.,  {Dolag} K.,  {Saro} A.,  {Bachmann} L.,  {Borgani} S.,
  {Burkert} A.,  2014, \mn@doi [\mnras] {10.1093/mnras/stu1023}, \href
  {https://ui.adsabs.harvard.edu/abs/2014MNRAS.442.2304H} {442, 2304}

\bibitem[\protect\citeauthoryear{{Hu{\v{s}}ko}, {Lacey}, {Schaye}, {Schaller}
  \& {Nobels}}{{Hu{\v{s}}ko} et~al.}{2022}]{Husko2022}
{Hu{\v{s}}ko} F.,  {Lacey} C.~G.,  {Schaye} J.,  {Schaller} M.,   {Nobels} F.
  S.~J.,  2022, \mn@doi [\mnras] {10.1093/mnras/stac2278}, \href
  {https://ui.adsabs.harvard.edu/abs/2022MNRAS.516.3750H} {516, 3750}

\bibitem[\protect\citeauthoryear{{Kaiser}}{{Kaiser}}{1986}]{Kaiser1986}
{Kaiser} N.,  1986, \mn@doi [\mnras] {10.1093/mnras/222.2.323}, \href
  {https://ui.adsabs.harvard.edu/abs/1986MNRAS.222..323K} {222, 323}

\bibitem[\protect\citeauthoryear{{Kaiser}}{{Kaiser}}{1991}]{Kaiser1991}
{Kaiser} N.,  1991, \mn@doi [\apj] {10.1086/170768}, \href
  {https://ui.adsabs.harvard.edu/abs/1991ApJ...383..104K} {383, 104}

\bibitem[\protect\citeauthoryear{{Kaviraj} et~al.,}{{Kaviraj}
  et~al.}{2017}]{HorizonAGN2017}
{Kaviraj} S.,  et~al., 2017, \mn@doi [\mnras] {10.1093/mnras/stx126}, \href
  {https://ui.adsabs.harvard.edu/abs/2017MNRAS.467.4739K} {467, 4739}

\bibitem[\protect\citeauthoryear{{Kugel} et~al.,}{{Kugel}
  et~al.}{2023}]{Kugel2023}
{Kugel} R.,  et~al., 2023, \mn@doi [\mnras] {10.1093/mnras/stad2540}, \href
  {https://ui.adsabs.harvard.edu/abs/2023MNRAS.526.6103K} {526, 6103}

\bibitem[\protect\citeauthoryear{{Lovisari} et~al.,}{{Lovisari}
  et~al.}{2017}]{Lovisari2017}
{Lovisari} L.,  et~al., 2017, \mn@doi [\apj] {10.3847/1538-4357/aa855f}, \href
  {https://ui.adsabs.harvard.edu/abs/2017ApJ...846...51L} {846, 51}

\bibitem[\protect\citeauthoryear{{Mantz}}{{Mantz}}{2019}]{Mantz2019}
{Mantz} A.~B.,  2019, \mn@doi [\mnras] {10.1093/mnras/stz320}, \href
  {https://ui.adsabs.harvard.edu/abs/2019MNRAS.485.4863M} {485, 4863}

\bibitem[\protect\citeauthoryear{Marini et~al.,}{Marini
  et~al.}{2024}]{marini2024detecting}
Marini I.,  et~al., 2024, Detecting Galaxy Groups and AGNs populating the local
  Universe in the eROSITA era (\mn@eprint {arXiv} {2404.12719})

\bibitem[\protect\citeauthoryear{{McCarthy}, {Schaye}, {Bird}  \& {Le
  Brun}}{{McCarthy} et~al.}{2017}]{Bahamas2017}
{McCarthy} I.~G.,  {Schaye} J.,  {Bird} S.,   {Le Brun} A. M.~C.,  2017,
  \mn@doi [\mnras] {10.1093/mnras/stw2792}, \href
  {https://ui.adsabs.harvard.edu/abs/2017MNRAS.465.2936M} {465, 2936}

\bibitem[\protect\citeauthoryear{{Miyatake} et~al.,}{{Miyatake}
  et~al.}{2023}]{HSC2023}
{Miyatake} H.,  et~al., 2023, \mn@doi [arXiv e-prints]
  {10.48550/arXiv.2304.00704}, \href
  {https://ui.adsabs.harvard.edu/abs/2023arXiv230400704M} {p. arXiv:2304.00704}

\bibitem[\protect\citeauthoryear{{Nelson}, {Pillepich}, {Ayromlou}, {Lee},
  {Lehle}, {Rohr}  \& {Truong}}{{Nelson} et~al.}{2023}]{ClusterTNG2023}
{Nelson} D.,  {Pillepich} A.,  {Ayromlou} M.,  {Lee} W.,  {Lehle} K.,  {Rohr}
  E.,   {Truong} N.,  2023, \mn@doi [arXiv e-prints]
  {10.48550/arXiv.2311.06338}, \href
  {https://ui.adsabs.harvard.edu/abs/2023arXiv231106338N} {p. arXiv:2311.06338}

\bibitem[\protect\citeauthoryear{{Ota} et~al.,}{{Ota} et~al.}{2023}]{Ota2023}
{Ota} N.,  et~al., 2023, \mn@doi [\aap] {10.1051/0004-6361/202244260}, \href
  {https://ui.adsabs.harvard.edu/abs/2023A&A...669A.110O} {669, A110}

\bibitem[\protect\citeauthoryear{{Pacaud} et~al.,}{{Pacaud}
  et~al.}{2018}]{XXL2018}
{Pacaud} F.,  et~al., 2018, \mn@doi [\aap] {10.1051/0004-6361/201834022}, \href
  {https://ui.adsabs.harvard.edu/abs/2018A&A...620A..10P} {620, A10}

\bibitem[\protect\citeauthoryear{{Pakmor} et~al.,}{{Pakmor}
  et~al.}{2023}]{MTNG2022}
{Pakmor} R.,  et~al., 2023, \mn@doi [\mnras] {10.1093/mnras/stac3620}, \href
  {https://ui.adsabs.harvard.edu/abs/2023MNRAS.524.2539P} {524, 2539}

\bibitem[\protect\citeauthoryear{{Pellissier}, {Hahn}  \&
  {Ferrari}}{{Pellissier} et~al.}{2023}]{Rhapsody2023}
{Pellissier} A.,  {Hahn} O.,   {Ferrari} C.,  2023, \mn@doi [\mnras]
  {10.1093/mnras/stad888}, \href
  {https://ui.adsabs.harvard.edu/abs/2023MNRAS.522..721P} {522, 721}

\bibitem[\protect\citeauthoryear{{Pillepich} et~al.,}{{Pillepich}
  et~al.}{2018}]{TNG2018}
{Pillepich} A.,  et~al., 2018, \mn@doi [\mnras] {10.1093/mnras/stx2656}, \href
  {https://ui.adsabs.harvard.edu/abs/2018MNRAS.473.4077P} {473, 4077}

\bibitem[\protect\citeauthoryear{{Planck Collaboration} et~al.,}{{Planck
  Collaboration} et~al.}{2016a}]{PlanckClustercosmo2016}
{Planck Collaboration} et~al., 2016a, \mn@doi [\aap]
  {10.1051/0004-6361/201525833}, \href
  {https://ui.adsabs.harvard.edu/abs/2016A&A...594A..24P} {594, A24}

\bibitem[\protect\citeauthoryear{{Planck Collaboration} et~al.,}{{Planck
  Collaboration} et~al.}{2016b}]{PLanckSZ2016}
{Planck Collaboration} et~al., 2016b, \mn@doi [\aap]
  {10.1051/0004-6361/201525823}, \href
  {https://ui.adsabs.harvard.edu/abs/2016A&A...594A..27P} {594, A27}

\bibitem[\protect\citeauthoryear{{Planck Collaboration} et~al.,}{{Planck
  Collaboration} et~al.}{2020}]{Planck2020}
{Planck Collaboration} et~al., 2020, \mn@doi [\aap]
  {10.1051/0004-6361/201833910}, \href
  {https://ui.adsabs.harvard.edu/abs/2020A&A...641A...6P} {641, A6}

\bibitem[\protect\citeauthoryear{{Ploeckinger} \& {Schaye}}{{Ploeckinger} \&
  {Schaye}}{2020}]{Ploeckinger2020}
{Ploeckinger} S.,  {Schaye} J.,  2020, \mn@doi [\mnras]
  {10.1093/mnras/staa2172}, \href
  {https://ui.adsabs.harvard.edu/abs/2020MNRAS.497.4857P} {497, 4857}

\bibitem[\protect\citeauthoryear{{Ramos-Ceja} et~al.,}{{Ramos-Ceja}
  et~al.}{2022}]{RamosCeja2022}
{Ramos-Ceja} M.~E.,  et~al., 2022, \mn@doi [\aap]
  {10.1051/0004-6361/202142214}, \href
  {https://ui.adsabs.harvard.edu/abs/2022A&A...661A..14R} {661, A14}

\bibitem[\protect\citeauthoryear{{Riess} et~al.,}{{Riess}
  et~al.}{2022}]{Ries2022}
{Riess} A.~G.,  et~al., 2022, \mn@doi [\apjl] {10.3847/2041-8213/ac5c5b}, \href
  {https://ui.adsabs.harvard.edu/abs/2022ApJ...934L...7R} {934, L7}

\bibitem[\protect\citeauthoryear{{Rossetti}, {Gastaldello}, {Eckert}, {Della
  Torre}, {Pantiri}, {Cazzoletti}  \& {Molendi}}{{Rossetti}
  et~al.}{2017}]{Rossetti2017}
{Rossetti} M.,  {Gastaldello} F.,  {Eckert} D.,  {Della Torre} M.,  {Pantiri}
  G.,  {Cazzoletti} P.,   {Molendi} S.,  2017, \mn@doi [\mnras]
  {10.1093/mnras/stx493}, \href
  {https://ui.adsabs.harvard.edu/abs/2017MNRAS.468.1917R} {468, 1917}

\bibitem[\protect\citeauthoryear{{Rozo}, {Bartlett}, {Evrard}  \&
  {Rykoff}}{{Rozo} et~al.}{2014}]{Rozo2014}
{Rozo} E.,  {Bartlett} J.~G.,  {Evrard} A.~E.,   {Rykoff} E.~S.,  2014, \mn@doi
  [\mnras] {10.1093/mnras/stt2161}, \href
  {https://ui.adsabs.harvard.edu/abs/2014MNRAS.438...78R} {438, 78}

\bibitem[\protect\citeauthoryear{{Rykoff} et~al.,}{{Rykoff}
  et~al.}{2014}]{Redmapper2014}
{Rykoff} E.~S.,  et~al., 2014, \mn@doi [\apj] {10.1088/0004-637X/785/2/104},
  \href {https://ui.adsabs.harvard.edu/abs/2014ApJ...785..104R} {785, 104}

\bibitem[\protect\citeauthoryear{{Rykoff} et~al.,}{{Rykoff}
  et~al.}{2016}]{Redmapper2016}
{Rykoff} E.~S.,  et~al., 2016, \mn@doi [\apjs] {10.3847/0067-0049/224/1/1},
  \href {https://ui.adsabs.harvard.edu/abs/2016ApJS..224....1R} {224, 1}

\bibitem[\protect\citeauthoryear{{Salcido}, {McCarthy}, {Kwan}, {Upadhye}  \&
  {Font}}{{Salcido} et~al.}{2023}]{Jaime2023}
{Salcido} J.,  {McCarthy} I.~G.,  {Kwan} J.,  {Upadhye} A.,   {Font} A.~S.,
  2023, \mn@doi [\mnras] {10.1093/mnras/stad1474}, \href
  {https://ui.adsabs.harvard.edu/abs/2023MNRAS.523.2247S} {523, 2247}

\bibitem[\protect\citeauthoryear{{Schaller} et~al.,}{{Schaller}
  et~al.}{2024}]{SWIFT2023}
{Schaller} M.,  et~al., 2024, \mn@doi [\mnras] {10.1093/mnras/stae922}, \href
  {https://ui.adsabs.harvard.edu/abs/2024MNRAS.530.2378S} {530, 2378}

\bibitem[\protect\citeauthoryear{{Schaye} \& {Dalla Vecchia}}{{Schaye} \&
  {Dalla Vecchia}}{2008}]{SchayeDV2008}
{Schaye} J.,  {Dalla Vecchia} C.,  2008, \mn@doi [\mnras]
  {10.1111/j.1365-2966.2007.12639.x}, \href
  {https://ui.adsabs.harvard.edu/abs/2008MNRAS.383.1210S} {383, 1210}

\bibitem[\protect\citeauthoryear{{Schaye} et~al.,}{{Schaye}
  et~al.}{2015}]{Eagle2015}
{Schaye} J.,  et~al., 2015, \mn@doi [\mnras] {10.1093/mnras/stu2058}, \href
  {https://ui.adsabs.harvard.edu/abs/2015MNRAS.446..521S} {446, 521}

\bibitem[\protect\citeauthoryear{{Schaye} et~al.,}{{Schaye}
  et~al.}{2023}]{FLAMINGOmain}
{Schaye} J.,  et~al., 2023, \mn@doi [\mnras] {10.1093/mnras/stad2419}, \href
  {https://ui.adsabs.harvard.edu/abs/2023MNRAS.526.4978S} {526, 4978}

\bibitem[\protect\citeauthoryear{{Springel}, {Di Matteo}  \&
  {Hernquist}}{{Springel} et~al.}{2005}]{Springel2005a}
{Springel} V.,  {Di Matteo} T.,   {Hernquist} L.,  2005, \mn@doi [\mnras]
  {10.1111/j.1365-2966.2005.09238.x}, \href
  {https://ui.adsabs.harvard.edu/abs/2005MNRAS.361..776S} {361, 776}

\bibitem[\protect\citeauthoryear{{Tinker}, {Kravtsov}, {Klypin}, {Abazajian},
  {Warren}, {Yepes}, {Gottl{\"o}ber}  \& {Holz}}{{Tinker}
  et~al.}{2008}]{Tinker2008}
{Tinker} J.,  {Kravtsov} A.~V.,  {Klypin} A.,  {Abazajian} K.,  {Warren} M.,
  {Yepes} G.,  {Gottl{\"o}ber} S.,   {Holz} D.~E.,  2008, \mn@doi [\apj]
  {10.1086/591439}, \href
  {https://ui.adsabs.harvard.edu/abs/2008ApJ...688..709T} {688, 709}

\bibitem[\protect\citeauthoryear{{Upsdell} et~al.,}{{Upsdell}
  et~al.}{2023}]{Upsdell2023}
{Upsdell} E.~W.,  et~al., 2023, \mn@doi [\mnras] {10.1093/mnras/stad1220},
  \href {https://ui.adsabs.harvard.edu/abs/2023MNRAS.522.5267U} {522, 5267}

\bibitem[\protect\citeauthoryear{Wendland}{Wendland}{1995}]{wendland1995}
Wendland H.,  1995, Advances in Computational Mathematics, 4, 389

\bibitem[\protect\citeauthoryear{{Wiersma}, {Schaye}, {Theuns}, {Dalla Vecchia}
   \& {Tornatore}}{{Wiersma} et~al.}{2009}]{Wiersma2009chemistry}
{Wiersma} R. P.~C.,  {Schaye} J.,  {Theuns} T.,  {Dalla Vecchia} C.,
  {Tornatore} L.,  2009, \mn@doi [\mnras] {10.1111/j.1365-2966.2009.15331.x},
  \href {https://ui.adsabs.harvard.edu/abs/2009MNRAS.399..574W} {399, 574}

\bibitem[\protect\citeauthoryear{{Willis} et~al.,}{{Willis}
  et~al.}{2021}]{Willis2021}
{Willis} J.~P.,  et~al., 2021, \mn@doi [\mnras] {10.1093/mnras/stab873}, \href
  {https://ui.adsabs.harvard.edu/abs/2021MNRAS.503.5624W} {503, 5624}

\makeatother
\end{thebibliography}




\appendix

\section{Fits at different redshifts}\label{sec:morez}
Tables~\ref{tab:fits0} - \ref{tab:fits3} contain the fits similar to those in Section~\ref{sec:lognorm} at the four other redshifts considered in this work. For each redshift, the general trends are similar to those found at $z=0.3$. We note that the fits for richness should be considered with care as they are not converged with the simulation resolution and, for most redshifts, the mass bins $M_{\rm 500c}>10^{14.0}~\rm{M}_{\odot}$ have a mean richness that is above ten. As the mean richness is very close to 10 for the $10^{13.5}~\rm{M_{\odot}}$ mass bin, going to that mass or lower will likely lead to results that suffer from small-number statistics. We omit the highest mass bin at $z=2$ as there are not enough high mass halos in the simulation volume to characterize the distribution.
\begin{table}
    \centering
        \caption{Values for the fits to Eq.~\ref{eq:fiteq} at $z=0$. The top four rows are for selection based on X-ray luminosity, the middle four for integrated Compton-Y, and the bottom for for galaxy richness. Note that for richness we only fit a lognormal, so we do not the include the parameters for the power-law tail.}
\begin{tabular}{c|c|l|l|l|l|l}
\hline
$a$ & $M_{\rm 500c}[\rm{M}_{\odot}]$ & $A$ & $\mu$ & $\sigma$ & $\log_{10}a_t$ & $\alpha$ \\
\hline
X-ray & $10^{13.0}~\rm{M}_{\odot}$ & $1.810\times10^{-3}$ & 40.8 & 0.37 & 41.5 & 2.06 \\
X-ray & $10^{13.5}~\rm{M}_{\odot}$ & $2.675\times10^{-3}$ & 42.0 & 0.24 & 42.2 & 3.57 \\
X-ray & $10^{14.0}~\rm{M}_{\odot}$ & $3.487\times10^{-3}$ & 42.9 & 0.19 & 43.2 & 4.23 \\
X-ray & $10^{14.5}~\rm{M}_{\odot}$ & $4.481\times10^{-3}$ & 43.8 & 0.15 & 43.9 & 6.11 \\
SZ & $10^{13.0}~\rm{M}_{\odot}$ & $3.737\times10^{-3}$ & -7.18 & 0.24 & -7.03 & 2.14 \\
SZ & $10^{13.5}~\rm{M}_{\odot}$ & $5.608\times10^{-3}$ & -6.15 & 0.17 & -6.05 & 3.47 \\
SZ & $10^{14.0}~\rm{M}_{\odot}$ & $7.220\times10^{-3}$ & -5.29 & 0.13 & -5.21 & 4.45 \\
SZ & $10^{14.5}~\rm{M}_{\odot}$ & $8.317\times10^{-3}$ & -4.48 & 0.11 & -4.40 & 5.38 \\
$\lambda$ & $10^{13.0}~\rm{M}_{\odot}$ & $3.213\times10^{-2}$ & 0.30 & 0.33 & - & - \\
$\lambda$ & $10^{13.5}~\rm{M}_{\odot}$ & $2.050\times10^{-2}$ & 0.81 & 0.30 & - & - \\
$\lambda$ & $10^{14.0}~\rm{M}_{\odot}$ & $3.347\times10^{-2}$ & 1.24 & 0.18 & - & - \\
$\lambda$ & $10^{14.5}~\rm{M}_{\odot}$ & $4.317\times10^{-2}$ & 1.70 & 0.14 & - & - \\
\hline
\end{tabular}

    \label{tab:fits0}
\end{table}

\begin{table}
    \centering
        \caption{As Table~\ref{tab:fits0}, but for $z=0.5$.}
\begin{tabular}{c|c|l|l|l|l|l}
\hline
$a$ & $M_{\rm 500c}[\rm{M}_{\odot}]$ & $A$ & $\mu$ & $\sigma$ & $\log_{10}a_t$ & $\alpha$ \\
\hline
X-ray & $10^{13.0}~\rm{M}_{\odot}$ & $1.923\times10^{-3}$ & 41.3 & -0.33 & 41.7 & 1.91 \\
X-ray & $10^{13.5}~\rm{M}_{\odot}$ & $2.977\times10^{-3}$ & 42.4 & -0.22 & 42.8 & 3.11 \\
X-ray & $10^{14.0}~\rm{M}_{\odot}$ & $4.380\times10^{-3}$ & 43.3 & 0.15 & 43.6 & 2.45 \\
X-ray & $10^{14.5}~\rm{M}_{\odot}$ & $4.395\times10^{-3}$ & 44.1 & -0.16 & 44.9 & 6.50 \\
SZ & $10^{13.0}~\rm{M}_{\odot}$ & $4.019\times10^{-3}$ & -6.98 & 0.22 & -6.86 & 2.12 \\
SZ & $10^{13.5}~\rm{M}_{\odot}$ & $5.475\times10^{-3}$ & -6.05 & 0.17 & -5.94 & 3.26 \\
SZ & $10^{14.0}~\rm{M}_{\odot}$ & $6.976\times10^{-3}$ & -5.23 & 0.13 & -5.14 & 4.20 \\
SZ & $10^{14.5}~\rm{M}_{\odot}$ & $7.975\times10^{-3}$ & -4.44 & 0.13 & -4.26 & 5.07 \\
$\lambda$ & $10^{13.0}~\rm{M}_{\odot}$ & $2.506\times10^{-2}$ & 0.44 & -0.34 & - & - \\
$\lambda$ & $10^{13.5}~\rm{M}_{\odot}$ & $2.540\times10^{-2}$ & 0.95 & 0.23 & - & - \\
$\lambda$ & $10^{14.0}~\rm{M}_{\odot}$ & $3.588\times10^{-2}$ & 1.34 & 0.17 & - & - \\
$\lambda$ & $10^{14.5}~\rm{M}_{\odot}$ & $4.709\times10^{-2}$ & 1.79 & 0.13 & - & - \\
\hline
\end{tabular}

    \label{tab:fits1}
\end{table}

\begin{table}
    \centering
        \caption{As Table~\ref{tab:fits0}, but for $z=1$.}
\begin{tabular}{c|c|l|l|l|l|l}
\hline
$a$ & $M_{\rm 500c}[\rm{M}_{\odot}]$ & $A$ & $\mu$ & $\sigma$ & $\log_{10}a_t$ & $\alpha$ \\
\hline
X-ray & $10^{13.0}~\rm{M}_{\odot}$ & $2.036\times10^{-3}$ & 41.8 & 0.30 & 42.0 & 1.99 \\
X-ray & $10^{13.5}~\rm{M}_{\odot}$ & $3.218\times10^{-3}$ & 42.8 & 0.20 & 43.1 & 2.80 \\
X-ray & $10^{14.0}~\rm{M}_{\odot}$ & $4.727\times10^{-3}$ & 43.7 & 0.14 & 43.9 & 3.39 \\
X-ray & $10^{14.5}~\rm{M}_{\odot}$ & $4.923\times10^{-3}$ & 44.4 & 0.14 & 44.9 & 6.50 \\
SZ & $10^{13.0}~\rm{M}_{\odot}$ & $4.428\times10^{-3}$ & -6.86 & 0.20 & -6.75 & 2.38 \\
SZ & $10^{13.5}~\rm{M}_{\odot}$ & $5.701\times10^{-3}$ & -5.99 & 0.16 & -5.88 & 3.44 \\
SZ & $10^{14.0}~\rm{M}_{\odot}$ & $7.126\times10^{-3}$ & -5.19 & 0.13 & -5.10 & 4.49 \\
SZ & $10^{14.5}~\rm{M}_{\odot}$ & $8.161\times10^{-3}$ & -4.41 & 0.12 & -4.32 & 6.12 \\
$\lambda$ & $10^{13.0}~\rm{M}_{\odot}$ & $2.324\times10^{-2}$ & 0.52 & -0.33 & - & - \\
$\lambda$ & $10^{13.5}~\rm{M}_{\odot}$ & $2.814\times10^{-2}$ & 0.99 & 0.21 & - & - \\
$\lambda$ & $10^{14.0}~\rm{M}_{\odot}$ & $3.922\times10^{-2}$ & 1.36 & 0.15 & - & - \\
$\lambda$ & $10^{14.5}~\rm{M}_{\odot}$ & $4.695\times10^{-2}$ & 1.80 & 0.13 & - & - \\
\hline
\end{tabular}
    \label{tab:fits2}
\end{table}

\begin{table}
    \centering
        \caption{As Table~\ref{tab:fits0}, but for $z=2$. Note that different from the other tables, we do not include the highest mass bin as there are insufficient halos to characterize the distributions at $z=2$}
\begin{tabular}{c|c|l|l|l|l|l}
\hline
$a$ & $M_{\rm 500c}[\rm{M}_{\odot}]$ & $A$ & $\mu$ & $\sigma$ & $\log_{10}a_t$ & $\alpha$ \\
\hline
X-ray & $10^{13.0}~\rm{M}_{\odot}$ & $2.217\times10^{-3}$ & 42.5 & -0.27 & 42.7 & 2.00 \\
X-ray & $10^{13.5}~\rm{M}_{\odot}$ & $3.330\times10^{-3}$ & 43.4 & 0.19 & 43.7 & 2.82 \\
X-ray & $10^{14.0}~\rm{M}_{\odot}$ & $4.190\times10^{-3}$ & 44.3 & 0.16 & 44.5 & 4.45 \\
SZ & $10^{13.0}~\rm{M}_{\odot}$ & $4.923\times10^{-3}$ & -6.76 & 0.19 & -6.62 & 3.11 \\
SZ & $10^{13.5}~\rm{M}_{\odot}$ & $6.225\times10^{-3}$ & -5.93 & 0.16 & -5.80 & 4.39 \\
SZ & $10^{14.0}~\rm{M}_{\odot}$ & $8.234\times10^{-3}$ & -5.14 & 0.13 & -5.04 & 6.94 \\
$\lambda$ & $10^{13.0}~\rm{M}_{\odot}$ & $2.557\times10^{-2}$ & 0.54 & -0.29 & - & - \\
$\lambda$ & $10^{13.5}~\rm{M}_{\odot}$ & $3.074\times10^{-2}$ & 0.97 & 0.19 & - & - \\
$\lambda$ & $10^{14.0}~\rm{M}_{\odot}$ & $4.414\times10^{-2}$ & 1.33 & 0.14 & - & - \\
\hline
\end{tabular}
    \label{tab:fits3}
\end{table}

\section{Using Compton-Y within $R_{\rm 500c}$}\label{sec:cyr500}
\begin{figure}
    \centering
    \includegraphics[width=\columnwidth]{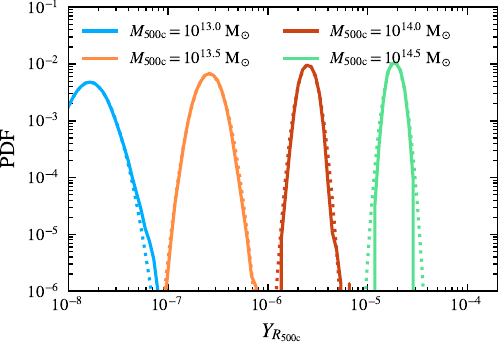}
    \caption{As the middle panel of Fig.~\ref{fig:lognorm}, but for an aperture of $R_\text{500c}$ instead of $5R_\text{500c}$. The shape of the distribution is very close to lognormal for every mass bin.}
    \label{fig:r5dists}
\end{figure}

For Compton-Y we use a fiducial aperture of $5R_\text{500c}$, which is motivated by \citet{PLanckSZ2016}, but is much larger than the apertures of $R_\text{500c}$ that we use for X-ray luminosity. In this section we investigate how the results for Compton-Y change if we use the same aperture as for the other observables. Fig.~\ref{fig:r5dists} shows the distribution of Compton-Y using a $R_{\rm 500c}$ aperture, akin to Fig.~\ref{fig:lognorm}. The different colours indicate different $0.1$~dex wide mass bins. The dotted lines show the results of fitting a lognormal distribution to each mass bin. The values of the fitted distribution can be found in Table~\ref{tab:r5fit}. Compared to the results for the $5R_{\rm 500c}$ aperture, the distributions now longer show a prominent power-law tail towards higher values of $Y_{\rm 500c}$. Furthermore, the distributions no longer overlap. Hence, the number of lower mass haloes that upscatter will be reduced.

\begin{figure}
    \centering
    \includegraphics[width=\columnwidth]{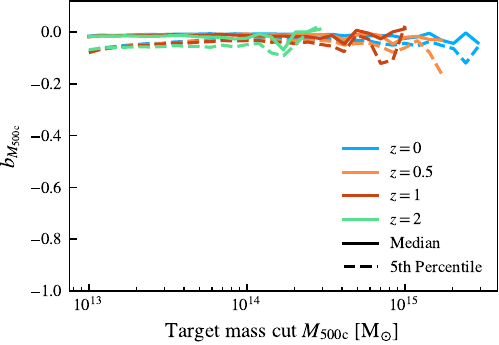}
    \caption{As the middle panel of Fig.~\ref{fig:changing mass}, but for an aperture of $R_\text{500c}$ instead of $5R_\text{500c}$. The sample mass bias is close to zero across all masses and redshifts.}
    \label{fig:bias_r5}
\end{figure}

In Fig.~\ref{fig:bias_r5} we show the sample mass bias at four redshifts indicated with different colours for a sample selected on Compton-Y within $R_{\rm 500c}$, akin to Fig.~\ref{fig:changing mass}. As expected from the previous figure, the smaller aperture leads to a a nearly mass-selected sample and the sample mass bias is close to zero for all target masses. Only the 5th percentile shows a slight ($\approx 5$ per cent) bias.

\begin{table}
\caption{Values for the fitting the functional form $f(x)=A\exp\left[-\frac{\left(\log_{10}\mu - x\right)^2}{\sigma^2}\right]$ to the distribution of Compton-Y within $R_{\rm 500c}$ for objects in a 0.1 dex width bin around the mass given in the first column. The second column gives the amplitude of the distribution, the third column the mean and the fourth column the scatter.}
\centering
\begin{tabular}{l|r|r|r|}
\hline
Mass $M_{\rm 500c}$ $[\rm{M}_{\odot}]$ & Amplitude & $\log_{10}\mu$ & $\sigma$ \\ \hline
$10^{13}$   & $4.79\times10^{-3}$   & -7.79   & $2.12\times10^{-1}$          \\
$10^{13.5}$ & $6.86\times10^{-3}$   & -6.59   & $1.50\times10^{-1}$          \\
$10^{14}$   & $9.95\times10^{-3}$   & -5.60   & $1.05\times10^{-1}$          \\
$10^{14.5}$ & $1.10\times10^{-2}$   & -4.73   & $9.52\times10^{-2}$          \\
\end{tabular}
\label{tab:r5fit}
\end{table}

\section{Convergence with numerical resolution and simulation box size}\label{sec:convergence}
\begin{figure}
    \centering
    \includegraphics[width=\columnwidth]{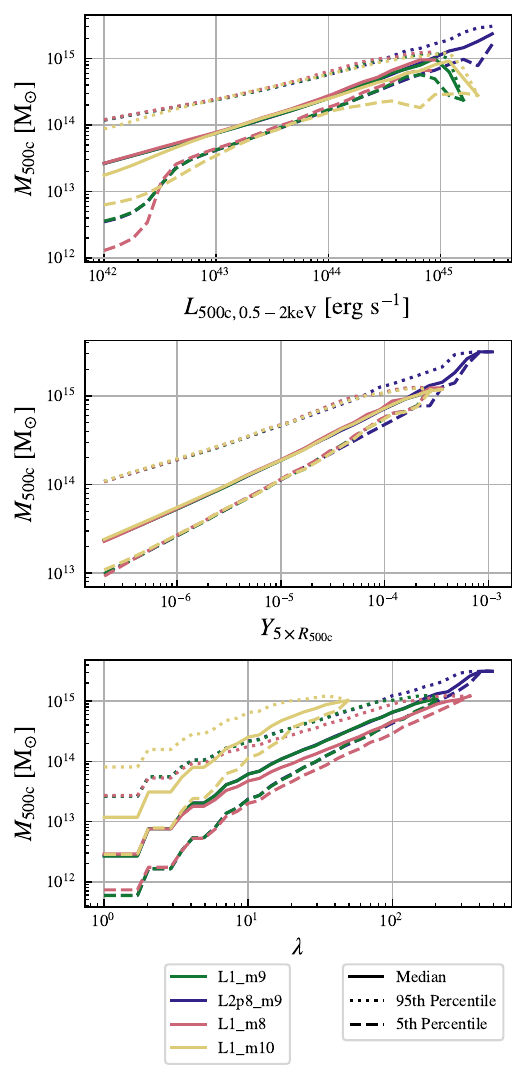}
    \caption{The solid lines show the median $M_{\rm 500c}$ for a sample selected using the cut on the quantity plotted along the x-axis. The different panels show the three different selection quantities, X-ray luminosity (top), thermal SZ Compton-Y (middle) and galaxy richness (bottom). The different colours show the results for the three different FLAMINGO resolutions and for the 1 and 2.8 $(\rm{Gpc})^3$ m9 boxes. The dashed (dotted) line indicates the 5th (95th) percentile mass for the sample after the cut. The mass distributions are converged with resolution for X-ray and SZ selection, but not for selection based on richness.}
    \label{fig:convergence_plot}
\end{figure}

In Fig.~\ref{fig:convergence_plot} we show the median (solid), 5th percentile (dashed) and 95th percentile (dotted) of $M_{\rm 500c}$ obtained for samples based on different selection cuts for the three different FLAMINGO resolutions, in a $(1~\text{Gpc})^3$ box at $z=0.3$. Additionally, we show the results for the $(2.8~\text{Gpc})^3$ box at intermediate resolution, see \cite{FLAMINGOmain} for the naming convention. Comparing L1\_m9 and L2p8\_m9 we find converged results for all but the largest halo masses, for which the sampling is much better in L2p8\_m9. The only box size effect is due to the improved statistics in a larger volume. For the SZ selection (middle panel) all percentiles are converged for all resolutions. 

For the X-ray luminosity selection (top panel) the median and 95th percentile are very close to being converged, with only the lowest resolution (m10) run decreasing slightly at the lowest masses. The largest difference is found for the 5th percentile. While the dip remains at roughly the same mass across the three resolutions, the 5th percentile drops more towards with higher resolution. This implies that the existence of the dip is not directly due to resolution effects, and could be caused due to an increase in scatter for haloes with this luminosity. The fact that the dip gets deeper with increasing resolution implies that at m9 resolution we do not yet resolve the full range of haloes that can up-scatter in our selection for the lowest luminosities. For our fiducial resolution (m9) the median is converged over the full target mass range and the 5th percentile is converged for target mass cuts of $M_\text{500c}\gtrsim 10^{14}\,\text{M}_\odot$.

\begin{figure}
    \centering
    \includegraphics[width=\columnwidth]{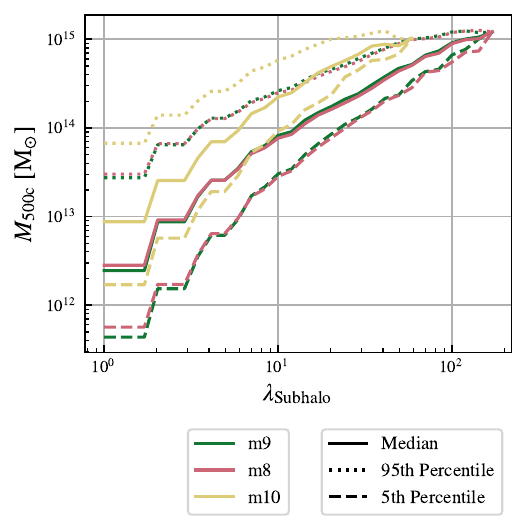}
    \caption{The solid lines show the median $M_{\rm 500c}$ for a sample selected using a cut on subhalo richness, defined using satellite subhaloes with a bound mass above $2\times10^{11}~\rm{M}_{\odot}$, and plotted along the horizontal axis. The different colours show the results for the three different FLAMINGO resolutions. The dashed (dotted) line indicates the 5th (95th) percentile mass for the sample after the cut. The halo mass distributions  for samples selected by cuts on subhalo richness are converged for m8 and m9, but not for m10.}
    \label{fig:Richness_convergence}
\end{figure}

For the richness selection we make use of both a cut in stellar mass and radius. As the stellar mass - halo mass relation is not converged at the high-mass end, see Fig.~9 from \citet{FLAMINGOmain}, it is not surprising that richness is not converged either. If we make a cut using the bound subhalo mass of the satellites instead of stellar mass, and pick a subhalo mass limit of $2\times10^{11}~\rm{M}_{\odot}$, which selects haloes close to the stellar mass limit of $10^{10.046}~\rm{M}_{\odot}$, then the m9 and m8 simulations do agree, as shown in Fig.~\ref{fig:Richness_convergence}. The subhalo mass cut is too low for the low resolution (m10) simulation to be converged, but the other two resolution simulations are converged if we select on subhalo richness. As the subhaloes are converged between m9 and m8, the differences we see for galaxy richness in the bottom panel of Figure~\ref{fig:convergence_plot} are caused by the differences in the stellar mass - halo mass relation. Both simulations match the stellar mass function up to $M_* = 10^{11.5}\,\text{M}_\odot$, so the fact that the richness selection is not fully converged is due to a combination of differences in the satellite fractions and imperfect calibration of the galaxy stellar mass function.

\bsp	
\label{lastpage}
\end{document}